# Shared-phase-dedicated-lane based intersection control with mixed traffic of human-driven vehicles and connected and automated vehicles

Wanjing Ma[1]

Tel.: (+86) 134-8223-6245

Email: mawanjing@tongji.edu.cn

Jinjue Li[1]

Tel: (+86) 176-0124-0717;

Email: 2011373@tongji.edu.cn

Chunhui Yu[1,*]

Tel.: (+86) 166-2118-2873

Email: hughyu90@tongji.edu.cn

[1] The Key Laboratory of Road and Traffic Engineering of the Ministry of Education, Tongji University, 4800 Cao'an Road, Shanghai 201804, China

*Corresponding author.

**Abstract**

Connected and automated vehicles (CAVs) and human-driven vehicles (HVs) are expected to coexist in the near future. CAV-dedicated lanes and phases have been explored to handle the uncertainty in the driving behavior of HVs in the mixed traffic environment. However, CAV-dedicated phases could significantly sacrifice HV benefits. This study proposes a shared-phase-dedicated-lane (SPDL)-based traffic control model at isolated intersections under the mixed traffic environment. Left-turn and through CAVs share CAV-dedicated lanes and cross the intersection during the shared phases with HVs. A three-level optimization model is developed. At the upper level, a standard NEMA (National Electrical Manufacturers Association) ring barrier structure is used for the signal optimization and barrier durations are optimized by dynamic programming to minimize the total vehicle delay. At the middle level, phase sequence and phase durations are optimized by enumeration for the given barrier from the upper level and the minimum vehicle delay is fed to the upper level. At the lower level, CAV platooning in the buffer zone and trajectory planning in the passing zone are conducted based on the signal timings of the barrier from the middle level and the travel time of CAVs is fed to the middle level. A rolling-horizon scheme is further designed for the dynamical implementation of the proposed model with time-varying traffic conditions. Numerical studies validate the advantages of the SPDL-based control over the blue-phase based control in previous studies in terms of average vehicle delay and intersection capacity. Further, the SPDL-based model is extended to serve as an alternative approach without the buffer zone.

# 1. Introduction

Traffic congestion in urban road networks mostly occurs at intersections, and congestion at intersections results in environmental problems and travel delay. Intersection control was developed with the intent to substantially reduce intersection congestion. With the development of connected vehicle (CV) and connected and automated vehicle (CAV) technologies, the traffic environment of intersection control has evolved from the traditional one with only human-driven vehicles (HVs) into that with CVs and CAVs.

Traditional intersection control relies on infrastructure-based detectors for data collection. These control methods mainly fall into four categories, namely, stage-, group-, lane-, and time-slot-based approaches. In stage-based approaches, compatible traffic movements move together simultaneously, which is called a stage. Green time is then allocated to each stage (Webster, 1958; Wong et al., 1996). In group-based approaches, green time is directly allocated to individual traffic movement to avoid the waste of green time (Tang and Nakamura, 2011; Jin and Ma, 2015; Wong, 1997; Wong and Yang, 1999; Jin and Ma, 2014). In the lane-based method, lane markings can be optimized together with signal timings for further improvement of traffic operation (Wong and Wong, 2003; Wong et al., 2006). Different from the repeated signal cycles with the same phase structure, the time-slot-based model proposed a generalized cycle structure with sub-cycles which could have different phase structures in Yu et al. (2020). In this model, the number of sub-cycles in one signal cycle was optimized together with the phase sequence, cycle length, and green splits in each sub-cycle.

The emergence of CVs provides trajectory data for intersection control. Compared with traditional aggregated data from infrastructure-based detectors, CV trajectories provide more detailed vehicle-level information via vehicle-to-vehicle (V2V) and vehicle-to-infrastructure (V2I) communication and are expected to improve traffic operation at intersections. In one research area, signalization at intersections is conducted with the assumption that all vehicles are connected. Vehicle trajectory data (e.g., speeds, acceleration rates, and location) are collected to estimate/predict vehicle arrival patterns at intersections and traffic condition (e.g., travel time and queue lengths) for signal timing (Goodall et al., 2013). In the other research area, different penetration rates of CVs are taken into consideration in intersection signalization. The challenge lies in the estimation/prediction of the states of unequipped vehicles (e.g., speeds and location) (Feng et al., 2015; Lioris et al., 2017) and traffic flow parameters (e.g., traffic volumes and shockwave speeds) (Ma et al., 2020; Tang et al., 2020), especially with low-resolution CV trajectory data. In addition, speed advisory for CVs has been explored at intersections (Bodenheimer et al., 2015). However, most studies assume that CVs could perfectly follow advisory speeds. Such scenarios are similar to CAV trajectory planning at intersections, which will be reviewed below.

Owing to the advances in CAV technology, the bilateral communication between intersections and vehicles facilitates the transfer of traffic information (e.g., intersection layouts and signal timings) from intersections to vehicles in real time for trajectory planning, and the collection of vehicle trajectory data (e.g., speeds and location) for signal control at intersections (Yu et al., 2018). Starting from the ideal traffic environment of a 100% CAV penetration rate, actuated, platoon-based, and planning-based intersection signal control have been proposed (Guo et al., 2019). Actuated control adjusts traffic signal timings according to real-time traffic demand (Li et al., 2014; Zhang et al., 2016). As traffic flow can be predicted based on CAV data, this method is also called adaptive traffic signal control. Platoon-based control groups vehicles into platoons and optimizes signal timings to enable platoons to pass through intersections without interruption (Liu et al., 2020). Planning-based control usually estimates every vehicle's actual travel time and predicts traffic conditions better (Li and Ban, 2017; Beak et al., 2017). To further improve traffic operation under a fully CAV environment, "signal-free" intersection control has been investigated (Levin and Rey, 2017; Stevanovic and Mitrovic, 2018;NOR and Namerikawa, 2020). Reservation- and optimization-based methods are applied to plan and control the trajectories of CAVs to prevent collisions at intersections without explicit signals.

However, the fully CAV environment cannot be realized in the near future. It is reported that the mixed traffic with human-driven vehicles (HVs) and CAVs will be prevalent in the next 20–30 years (Zheng et al., 2020). Several studies have focused on the trajectory planning for CAVs in the mixed traffic environment together with signal optimization at intersections when CAVs and HVs share lanes and signals. A commonly used approach is the formation of platoons with mixed CAVs and HVs by controlling the trajectories of CAVs as platoon-leading vehicles (Yao et al., 2020; Niroumand et al., 2020; Yang et al., 2021). The platoon-leading CAV trajectories and signal timings are optimized together for the improvement of traffic operation performance such as reducing vehicle delay and energy consumption and raising intersection capacity. Yao et al. (2020) proposed an optimization framework for CAV trajectory planning and traffic signal timing at an intersection under the mixed traffic environment. At the upper level, the optimization of the traffic signal timings was modeled as a dynamic programming (DP) problem to minimize vehicle delay. At the lower level, the trajectories of CAVs were optimized to form mixed platoons and lead them to cross the intersection for minimum gasoline consumption. To fully utilize the benefits of CAVs, Shunsuke and Raj (2019) investigated a decentralized intersection protocol for the application of the control approaches under the fully CAV environment to the mixed traffic environment. When there were no HVs around the intersection, CAVs synchronized and crossed the intersection ignoring traffic signals. Otherwise, they followed traffic signals as HVs did. Mixed traffic control at "signal-free" intersections has also been investigated (Zhang and Cassandras, 2019). When a CAV was not constrained by preceding HVs, it followed an optimal acceleration profile for minimum energy consumption with the consideration of collision avoidance. Otherwise, the CAV adaptively maintained a safe trailing distance from its preceding HV. HVs were controlled by conflict areas and priority rules (e.g., stop signs) defined in VISSIM.

One main difficulty in mixed traffic control is the uncertainties in the driving behavior of HVs (Zheng et al., 2020). These uncertainties will negatively influence the operational efficiency and safety of intersection control in a mixed traffic environment where HVs and CAVs share lanes. Therefore, Rey and Levin (2019) proposed dedicated phases and lanes for CAVs and then introduced a pressure-based decentralized hybrid-network control model to maximize the network throughput. The dedicated phases were called blue phases (BPs). During a BP, CAVs in each dedicated lane were discharged simultaneously. The sequence and times of CAVs passing conflict points within the intersection were controlled to avoid collisions. The hybrid network control strategy outperformed traditional green phase traffic signal control under high congestion conditions. However, if the proportion of CAVs was high in the network, the travel time of HVs might be considerably penalized (Rey and Levin, 2019). Moreover, only the arrival speeds and times of CAVs at the stop line were optimized with point-queues. CAV trajectory planning in approaching lanes was missing, which could further improve the traffic operation.

In view of these research gaps, this study proposes a shared-phase-dedicated-lane (SPDL)-based traffic control model at isolated intersections under a mixed traffic environment of HVs and CAVs. Dedicated lanes are shared by left-turn and through CAVs. CAVs and HVs share phases. A three-level optimization model is formulated. At the upper level, a standard NEMA (National Electrical Manufacturers Association) ring barrier structure is used for the signal optimization and barrier durations are optimized by DP to minimize the total vehicle delay. At the middle level, phase sequence and phase durations are optimized by enumeration for the given barrier from the upper level and the minimum vehicle delay is fed to the upper level. At the lower level, CAV platooning in the buffer zone and trajectory planning in the passing zone are conducted based on the signal timings of the barrier from the middle level and the travel time of CAVs is fed to the middle level. Furthermore, a rolling-horizon scheme is designed to implement the proposed model with varying traffic conditions.

This study makes the following contributions. 1) An intersection control model is proposed for managing mixed traffic with HVs and CAVs which combines traditional signal timing schemes, CAV-dedicated lanes, and CAV trajectory planning in approaching lanes. 2) Left-turn and through CAVs share CAV-dedicated lanes with trajectory planning and cross the intersection during the shared phases with HVs for better utilization of spatiotemporal resources. 3) The proposed intersection control framework is extended to remove the restriction of defining buffer zones for generalization.

The remainder of this paper is organized as follows. Section 2 presents the notations and problem description. Section 3 introduces the SPDL-based control model. Section 4 presents the rolling horizon scheme. Section 5 presents the numerical studies. Section 6 concludes the paper and discusses the directions for future research.

## 2. Notations and problem description

### 2.1. Notations

**Table 1** summarizes the notations of the parameters and the variables used in this study. The notations within brackets represent the same variable with subscripts omitted for simplicity.

**Table 1** Parameters and variables used in this study.

| General notations/parameters | |
|---|---|
| $j$ | Barrier group index |
| $r$ | Ring index |
| $p$ | Phase index in the cell of a ring and a barrier group ($p = 1,2$) |
| $m$ | HV movement ($m = 1,\ldots,8$). Denote movement sets $M_1 = \{1,2,3,4\}$ and $M_2 = \{5,6,7,8\}$ |
| $t_0$ | Time step at which the optimized signal plan begins |
| $\Delta t$ | Length of one time step, s |
| $\omega$ | CAV index |
| Parameters in upper level model | |

| $R_{p,r,j}$ | Green interval after phase $p$ in ring $r$ in barrier $j$ including yellow change and red clearance times, s |
|---|---|
| $G_{p,r,j}^{min}$ | Minimum green times of phase $p$ in ring $r$ in barrier group $j$, s |
| $G_{p,r,j}^{max}$ | Maximum green times of phase $p$ in ring $r$ in barrier group $j$, s |

Variables in upper level model

| $x_j$ | Time steps assigned to barrier group $j$ (i.e., the decision variable at stage $j$) |
|---|---|
| $s_j$ | Total time steps allocated up to barrier group $j-1$ (i.e., the state at stage $j$) |
| $X_j^{min}$ | Minimum time steps that could be assigned to barrier group $j$ |
| $X_j^{max}$ | Maximum time steps that could be assigned to barrier group $j$ |
| $f_j(s_j, x_j)$ | Performance function at stage $j$ given state $s_j$ and decision variable $x_j$, veh $\cdot$ s |
| $v_j(s_j)$ | Accumulated value function up to stage $j$ given state $s_j$, veh $\cdot$ s |

Parameters in middle level model

| $q_m^a(t)$ | Arrival flow rate of HVs with movement $m$, veh/h |
|---|---|
| $q_m^s$ | Saturation flow rate (in vehicles) of HVs with movement $m$, veh/h |
| $l_{p,r,j}^0$ | Initial queue length of HVs of phase $p$ in ring $r$ in barrier group $j$ at time step $t_0$, veh |
| $t_{free}^\omega$ | Free-flow travel time of vehicle $\omega$ to cross the intersection, s |

Variables in middle level model

| $d_j^h$ | HV delay in barrier group $j$, veh $\cdot$ s |
|---|---|
| $d_{p,r,j}^h$ | HV delay of phase $p$ in ring $r$ in barrier group $j$, veh $\cdot$ s |
| $d_j^c$ | CAV delay in barrier group $j$, veh $\cdot$ s |
| $l_{p,r,j}(t)$ | Queue length of HVs of phase $p$ in ring $r$ in barrier group $j$ at time step $t$, veh |
| $q_{p,r,j}^a(t)$ | Arrival flow rate of HVs of phase $p$ in ring $r$ in barrier group $j$ at time step $t$, veh/h |
| $q_{p,r,j}^d(t)$ | Departure flow rate of HVs of phase $p$ in ring $r$ in barrier group $j$ at time step $t$, veh/h |
| $q_{p,r,j}^s$ | Saturation flow rate of HVs of phase $p$ in ring $r$ in barrier group $j$, veh/h |
| $t_j^\omega$ | Actual travel time of vehicle $\omega$ in barrier group $j$, s |
| $\alpha_{p,r,j}^m$ | 1 if phase $p$ in ring $r$ selects movement $m$ in $M_r$, where $M_1 = \{1,2,3,4\}$ and $M_2 = \{5,6,7,8\}$. Denote $\boldsymbol{\alpha}_j$ as the vector of $\alpha_{p,r,j}^m$ indicating the phase sequence in barrier group $j$ |
| $g_{p,r,j}$ | Green time steps assigned to phase $p$ in ring $r$ in barrier group $j$. Denote $\boldsymbol{g}_j$ as the vector of $g_{p,r,j}$ indicating the phase duration in barrier group $j$ |

Parameters in lower level model

| | |
|---|---|
| $t_0^\omega$ | Entry time of CAV $\omega$ into the passing zone, s |
| $t_{p,r,j}^g$ | Beginning time of phase $p$ in ring $r$ in barrier group $j$, s |
| $\tau^\omega$ | Reaction time of CAV $\omega$, s |
| $v_0^\omega$ | Entry speed of CAV $\omega$ into the passing zone, m/s |
| $v_0^U$ | Upper boundary of the entry speed into the passing zone, m/s |
| $v_0^L$ | Lower boundary of the entry speed into the passing zone, m/s |
| $v_{max}$ | Free flow speed of CAVs, m/s |
| $a^L$ | Maximum deceleration, m/s² |
| $a^U$ | Maximum acceleration, m/s² |
| $L$ | Length of the passing zone, m |
| $l^\omega$ | Length of CAV $\omega$, m |
| $d_{jam}^\omega$ | Jam spacing between CAV $\omega$ and its preceding vehicle, m |

Variable in lower level model

| | |
|---|---|
| $a_i^\omega$ | Acceleration/deceleration rate of CAV $\omega$ at the $i^{th}$ trajectory segment ($i = 1,2,3$), m/s² |
| $x^\omega(t)$ | Travelled distance of CAV $\omega$ at time $t$, m |
| $v^\omega(t)$ | Speed of CAV $\omega$ at time $t$, m/s |
| $t_f^\omega$ | Time point when CAV $\omega$ passes the stop line, s |

*2.2. Problem description*

**Fig. 1** illustrates the layout of a typical isolated intersection with four arms. There are three vehicle movements (i.e., left-turn, through, and right-turn) in each arm. There is a dedicated lane for left-turn and through CAVs in each arm. Right-turn CAVs and HVs share the right-turn lanes and are not controlled by signals. CAVs and HVs share phases. Each arm is divided into a buffer zone and a passing zone. Arriving CAVs form platoons in the buffer zone and could wait for the entry time into the passing zone. CAV platoons then follow planned trajectories in the passing zone to cross the intersection. CAV platooning, the entry time of each CAV platoon, and CAV platoon trajectories in the passing zone are optimized in one framework together with signal timing plans including cycle length, phase sequence, and phase durations. CAV platooning serves two purposes: 1) to make through and left-turn CAVs share the dedicated lanes to cross the intersection during the phases of HVs; and 2) to simplify the trajectory optimization for CAVs. Only the trajectories of platoon-leading CAVs are optimized, which lead platoon-following CAVs to cross the intersection without stops at stop lines. For simplicity, the following assumptions are made:

1) The arrival information of CAVs and HVs within the prediction horizon $T$ can be collected or predicted.
2) HV queue length could be detected by infrastructure-based detectors, e.g., video cameras.
3) CAVs can finish platooning and HVs can finish lane-changing in the buffer zone.
4) CAV platoons leave the buffer zone and enter the passing zone at a speed between a lower bound $v_0^L$ and an upper bound $v_0^U$.

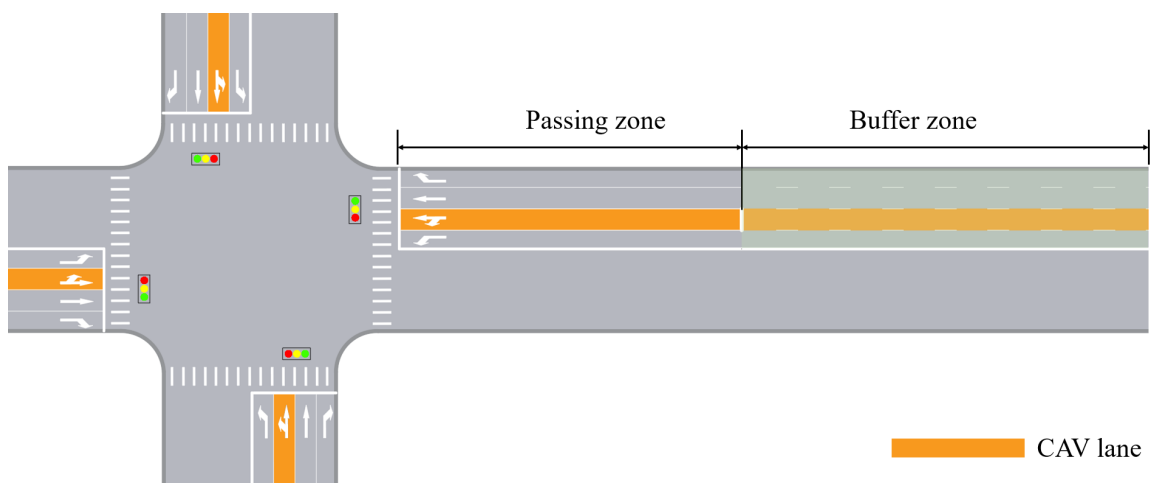

**Fig. 1** A typical intersection with dedicated lanes for CAVs.

## 3. Model formulation

### 3.1. Model framework of SPDL-based control

**Fig. 2** shows the model framework of the SPDL-based intersection control. The inputs to the control model are the arrival information of CAVs and HVs within the prediction horizon $T$. The CAV arrival information includes the arrival time at the buffer zone and the movements at the intersection, which could be collected by V2I communication in advance. The HV arrival information is the average arrival rate of each vehicle movement, which is simply predicted by previous HV arrival information in this study. A three-level optimization model is established to optimize signal timings and CAV trajectories in one framework. The optimization of barrier durations at the upper level is modeled as a DP problem to minimize the total delay of CAVs and HVs. Based on the barrier from the upper level model, the middle level model optimizes phase sequence and phase durations and the minimum vehicle delay is fed to the upper level. Based on the signal timings of the barrier from the middle level model, CAV platooning in the buffer zone and trajectory planning in the passing zone are conducted at the lower level, and the travel time of CAVs is fed to the middle level. The outputs of the control model include the optimal signal timings at the intersection and the planned CAV trajectories in the passing zone.

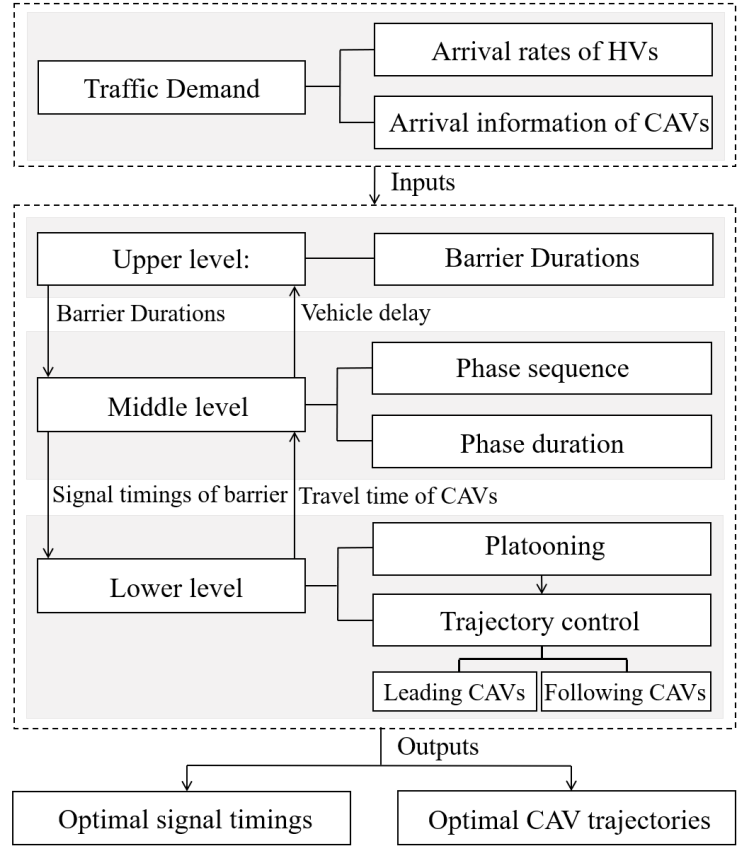

**Fig. 2** Framework of SPDL-based intersection control model.

*3.2. Upper level model*

A standard NEMA ring barrier structure is applied as shown in **Fig. 3**. The cycle length, phase sequence, and phase duration are optimized. For the convenience of modeling, the two phases corresponding to each ring and each barrier group are indexed by $p = 1, 2$. For example, the phase indexes of movements 1, 5, 3, and 7 are $p = 1$ in **Fig. 3**. The signal optimization at the upper level is modeled as a DP problem based on discrete time. Each barrier group is regarded as a stage, and one signal cycle is comprised of two stages. It is noted that the DP modeling approaches with a predetermined planning horizon (e.g., 80 s in Feng et al. (2015).) in existing studies may underestimate vehicle delay with over-saturated traffic. Because the delay of the vehicles that cannot be discharged in the planning horizon cannot be fully counted. Moreover, it is difficult to determine a feasible planning horizon for varying traffic demands. However, a sufficiently long planning horizon increases computational burden. To handle this issue, this study uses a repeated-stage structure as shown in **Fig. 4**. Stage 1 and stage 2 are repeated until the CAVs and HVs in the prediction horizon $T$ are all discharged. The benefits include: 1) the stage number could adaptively vary with different demand, which is the minimum number to discharge all vehicles; 2) the delay of vehicles with multiple stops could be fully taken into consideration; and 3) the delay estimation is simplified with the repeated-stage structure for the improvement of computational efficiency.

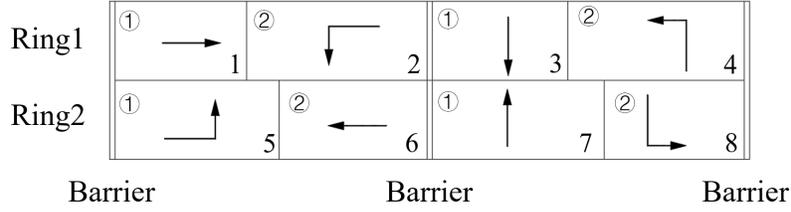

**Fig. 3** Ring barrier controller structure.

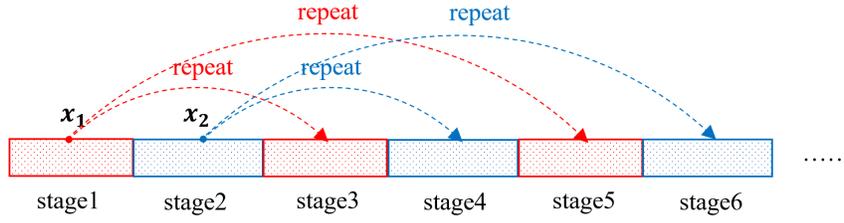

**Fig. 4** A repeated-stage structure in DP modeling.

The state $s_j$ at stage $j$ is the total number of time steps allocated up to stage $j-1$. The decision variable $x_j$ at stage $j$ is the time steps assigned to stage $j$. $x_j$ is bounded by the minimum and maximum time steps ($X_j^{min}$ and $X_j^{max}$):

$$X_j^{min} \leq x_j \leq X_j^{max} \tag{1}$$

$X_j^{min}$ and $X_j^{max}$ are determined by

$$X_j^{min} = \frac{\max\{G_{1,1,j}^{min}+R_{1,1,j}+G_{2,1,j}^{min}+R_{2,1,j}, G_{1,2,j}^{min}+R_{1,2,j}+G_{2,2,j}^{min}+R_{2,2,j}\}}{\Delta t} \tag{2}$$

$$X_j^{max} = \frac{\max\{G_{1,1,j}^{max}+R_{1,1,j}+G_{2,1,j}^{max}+R_{2,1,j}, G_{1,2,j}^{max}+R_{1,2,j}+G_{2,2,j}^{max}+R_{2,2,j}\}}{\Delta t} \tag{3}$$

where $G_{p,r,j}^{min}$ and $G_{p,r,j}^{max}$ are the minimum and maximum green times of phase $p$ in ring $r$ in barrier $j$; $R_{p,r,j}$ is the green interval after phase $p$ in ring $r$ in barrier $j$ including yellow change and red clearance times; $\Delta t$ is the length of one time step. $\Delta t$ should be properly selected to guarantee that $X_j^{min}$ and $X_j^{max}$ are integers. The state transition equation is

$$s_{j+1} = s_j + x_j \tag{4}$$

3.2.1. Forward recursion

The repeated-stage structure in **Fig. 4** makes it possible to calculate the total vehicle delay when the signal timings at the first two stage are determined. A forward recursion with three stages is designed for the DP algorithm:

**Step 1:** Initialize $j = 1$, $s_j = 0$, and $v_j(s_j) = 0$.

**Step 2:** Set $j = j + 1$.

**Step 3:** For $s_j = \sum_{j'=1}^{j-1} X_{j'}^{min}, \dots, \sum_{j'=1}^{j-1} X_{j'}^{max}$

$$v_j(s_j) = \min_{X_{j-1}^{min} \leq x_{j-1} \leq X_{j-1}^{max}} \{v_{j-1}(s_{j-1}) + f_{j-1}(s_{j-1}, x_{j-1})\}$$

Record $x_{j-1}^*(s_j)$ as the optimal solution.

**Step 4:** If $j < 3$, then go to **Step 2**. Otherwise, stop.

where $v_j(s_j)$ is the accumulated value function up to stage $j$ given state $s_j$; $f_j(s_j, x_j)$ is the performance function at stage $j$ given state $s_j$ and decision variable $x_j$. $f_j(s_j, x_j)$ is calculated in the middle level model by optimizing the phase sequence, phase duration, and CAV trajectories for minimum travel delay of CAVs and HVs.

3.2.2. Backward recursion

The optimal decision $x_j^*$ at stage 1 and stage 2 can be retrieved by a backward recursion:

**Step 1:** Set $j = 3$, $s_j^* = \mathrm{argmin}\, v_j(s_j)$

**Step 2:** Set $x_{j-1}^* = x_{j-1}^*(s_j^*)$

**Step 3:** Set $s_{j-1}^* = s_j^* - x_{j-1}^*$

**Step 4:** Set $j = j - 1$

**Step 5:** If $j = 1$, then stop and output the optimal solution $x_j^*$ together with the corresponding optimal phase sequence, phase duration, and planned CAV trajectories. Otherwise, go to **Step 2**.

### 3.3. Middle level model

Given state $s_j$ and decision variable $x_j$ at stage $j$ from the upper level model, the middle level model optimizes the phase sequence and phase durations of barrier group $j$ together with the trajectory planning for the CAVs. The minimum total travel delay is fed to the upper level model as the value of the performance function $f_j(s_j, x_j)$.

#### 3.3.1. Calculation of HV travel delay

The HV delay $d_j^h$ in barrier group $j$ is calculated as the total travel delay of HVs experienced during barrier group $j$. $d_j^h$ is formulated as

$$d_j^h = \sum_{r=1}^{2} \sum_{p=1}^{2} d_{p,r,j}^h \tag{5}$$

$$d_{p,r,j}^h = \sum_{t=t_0+s_j+1}^{t_0+s_j+x_j} l_{p,r,j}(t)\Delta t, \forall p = 1,2; r = 1,2 \tag{6}$$

$$l_{p,r,j}(t) = \max\{l_{p,r,j}(t-1) + q_{p,r,j}^a(t) - q_{p,r,j}^d(t), 0\}, \forall t = t_0 + s_j + 1, \ldots, t_0 + s_j + x_j; p = 1,2; r = 1,2 \tag{7}$$

$$l_{p,r,j}(t_0 + s_j) = \begin{cases} l_{p,r,j-1}(t_0 + s_{j-1} + x_{j-1}), & \text{if } j \geq 2 \\ l_{p,r,1}(t_0) = l_{p,r,1}^0, & \text{if } j = 1 \end{cases}, \forall p = 1,2; r = 1,2 \tag{8}$$

$$q_{1,r,j}^d(t) = \begin{cases} \min\{q_{1,r,j}^s, l_{1,r,j}(t-1) + q_{1,r,j}^a(t)\}, & \text{if } t_0 + s_j < t \leq t_0 + s_j + g_{1,r,j} \\ 0, & \text{otherwise} \end{cases}, \forall r = 1,2 \tag{9}$$

$$q_{2,r,j}^d(t) = \begin{cases} \min\{q_{2,r,j}^s, l_{2,r,j}(t-1) + q_{2,r,j}^a(t)\}, & \text{if } t_0 + s_j + g_{1,r,j} + \frac{R_{1,r,j}}{\Delta t} < t \leq t_0 + s_j + x_j - \frac{R_{2,r,j}}{\Delta t} \\ 0, & \text{otherwise} \end{cases}, \forall r = 1,2$$

$$\tag{10}$$

$$q_{p,r,j}^s = \sum_m \alpha_{p,r,j}^m q_m^s, \forall p = 1,2; r = 1,2 \tag{11}$$

$$q_{p,r,j}^a(t) = \sum_m \alpha_{p,r,j}^m q_m^a(t), \forall p = 1,2; r = 1,2 \tag{12}$$

where $d_{p,r,j}^h$ is the travel delay of HVs of phase $p$ in ring $r$ in barrier group $j$ that is experienced during barrier group $j$; $t_0$ is the time step at which the optimized signal plan begins; $l_{p,r,j}(t)$ is the queue length (in vehicles) of HVs of phase $p$ in ring $r$ in barrier group $j$ at time step $t$; $l_{p,r,j}^0$ is the initial queue length of HVs of phase $p$ in ring $r$ in barrier group $j$ at time step $t_0$. $q_{p,r,j}^a(t)$ is the arrival rate (in vehicles) of HVs of phase $p$ in ring $r$ in barrier group $j$ at time step $t$; $q_{p,r,j}^d(t)$ is the departure flow rate (in vehicles) of HVs of phase $p$ in ring $r$ in barrier group $j$ at time step $t$;

$q_{p,r,j}^s$ is the saturation flow rate (in vehicles) of HVs of phase $p$ in ring $r$ in barrier group $j$; $g_{p,r,j}$ is the time steps assigned to phase $p$ in ring $r$ in barrier group $j$; $R_{p,r,j}/\Delta t$ should be an integer; $q_m^s$ is the saturation flow rate (in vehicles) of HVs with movement $m$; $q_m^a(t)$ is the arrival flow rate (in vehicles) of HVs with movement $m$. $\alpha_{p,r,j}^m = 1$ if HVs with movement $m$ use phase $p$ in ring $r$ in barrier group $j$ to cross the intersection; and $\alpha_{p,r,j}^m = 0$, otherwise.

Eq. (5) calculates the total travel delay of all the phases in barrier group $j$. Eq. (6) calculates the travel delay of each phase in barrier group $j$. Eq. (7) describes the evolution of the queue length of HVs of phase $p$ in ring $r$ in barrier group $j$. Eq. (8) shows the queue length at the start of barrier group $j \geq 2$ equals that at the end of barrier group $j - 1$, which is determined in the planning at stage $j - 1$. Eq. (8) also indicates the initial queue length at time step $t_0$ when the optimized signal plan begins. Eqs. (9) and (10) calculate the departure rates of HVs of phase $p = 1$ and phase $p = 2$ in ring $r$ in barrier group $j$, respectively. Eq. (11) determines the saturation flow rate of HVs of phase $p$ in ring $r$ in barrier group $j$, which is related to the phase sequence. Eq. (12) determines the arrival flow rate of HVs of phase $p$ in ring $r$ in barrier group $j$, which is related to the phase sequence.

### 3.3.2. Calculation of CAV travel delay

The CAV delay $d_j^c$ in barrier group $j$ is calculated as the total travel delay of CAVs that could cross the intersection within barrier group $j$. $d_j^c$ is formulated as

$$d_j^c = \sum_{\omega \in \Omega}(t_j^\omega - t_{free}^\omega) \tag{13}$$

$$t_j^\omega = t^\omega(\boldsymbol{\alpha}_j, \boldsymbol{g}_j), \forall \omega \in \Omega \tag{14}$$

where $\Omega$ is the set of CAVs arriving at the intersection in the prediction horizon $T$; $t_{free}^\omega$ is the free-flow travel time of vehicle $\omega$ to cross the intersection; $t_j^\omega$ is the actual travel time of vehicle $\omega$ in barrier group $j$; $\boldsymbol{\alpha}_j$ is the vector of $\alpha_{p,r,j}^m$ indicating the phase sequence in barrier group $j$; $\boldsymbol{g}_j$ is the vector of $g_{p,r,j}$ indicating the phase duration in barrier group $j$. $t_j^\omega$ is related to the signal timings in barrier group $j$ and is determined by the CAV trajectory planning in the lower level model.

### 3.3.3. Signal constraints

The phase sequence and phase duration are both optimized in barrier group $j$. The signal constraints include Eqs.(15)-(22):

$$\sum_{m \in M_r} \alpha_{p,r,j}^m = 1, \forall p = 1,2; r = 1,2 \tag{15}$$

$$\sum_{p=1}^{2} \alpha_{p,r,j}^m \leq 1, \forall m \in M_r; r = 1,2 \tag{16}$$

$$\sum_{m=1}^{2} \alpha_{1,1,j}^m = \sum_{m=1}^{2} \alpha_{2,1,j}^m \tag{17}$$

$$\sum_{m=5}^{6} \alpha_{1,2,j}^m = \sum_{m=5}^{6} \alpha_{2,2,j}^m \tag{18}$$

$$\sum_{p=1}^{2} \sum_{m=1}^{2} \alpha_{p,1,j}^m = \sum_{p=1}^{2} \sum_{m=5}^{6} \alpha_{p,2,j}^m \tag{19}$$

$$\sum_{m=1}^{2} \alpha_{1,1,j}^m = \sum_{m=3}^{4} \alpha_{1,1,j-1}^m \tag{20}$$

$$G_{p,r,j}^{min} \leq g_{p,r,j} \Delta t \leq G_{p,r,j}^{max}, \forall p = 1,2; r = 1,2 \tag{21}$$

$$\sum_{p=1}^{2} (g_{p,r,j} + \frac{R_{p,r,j}}{\Delta t}) = x_j, \forall r = 1,2 \tag{22}$$

Eq. (15) indicates phase $p$ in ring $r$ selects movement $m$ in $M_r$, where $M_1 = \{1,2,3,4\}$ and $M_2 = \{5,6,7,8\}$. Eq. (16) indicates phase 1 and phase 2 in each ring cannot select the same movement. Eqs. (17) and (18) indicate phase 1 and phase 2 in each ring select north-/south-bound movements or east-/west-bound movements. Eq. (19) indicates the phases in ring 1 and ring 2 select compatible movements. Eq. (20) indicates the phase selection in adjacent barrier groups should be different together with Eqs. (17) and (18). Eq. (21) indicates the maximum and minimum green time constraints. And Eq. (22) indicates that the sum of the duration of the phases in each ring equals $x_j$. Note that $\alpha_{p,r,0}^m$ indicates the phase selection in the second barrier group in the current signal cycle, which is known.

3.3.4. Calculation of performance functions

Due to the repeated-stage structure, the calculation of $f_j(s_j, x_j)$ is different at stage $j = 1$ and stage $j = 2$. At stage $j = 1$, $f_1(s_1, x_1)$ is the sum of the travel delay of HVs experienced during barrier group $j = 1$ and the travel delay of CAVs that could cross the intersection within barrier group $j = 1$. $f_1(s_1, x_1)$ is determined by solving the following problem (**P1**):

(**P1**)

$$f_1(s_1, x_1) = \min d_1^h + d_1^c \tag{23}$$

Constraints include Eqs. (5)–(22) for $j = 1$.

At stage $j = 2$, $f_2(s_2, x_2)$ is the sum of the travel delay of HVs experienced during barrier groups $j = 2, \dots, J$ and the travel delay of CAVs that could cross the intersection within barrier groups $j = 2, \dots, J$. $J$ is the stage number to discharge all HVs and CAVs in the prediction horizon $T$. $J$ is determined by the predicted demand and the signal timings at the first two stages. $f_2(s_2, x_2)$ is determined by solving the following problem (**P2**):

(**P2**)

$$f_2(s_2, x_2) = \min \sum_{j=2}^{J} d_j^h + d_j^c \tag{24}$$

Constraints include Eqs. (5)–(22) for $j = 2, \dots, J$.

Enumeration is used to solve the mixed-integer nonlinear programming models (**P1**) and (**P2**). $f_1(s_1, x_1)$ and $f_2(s_2, x_2)$ are fed to the upper level model to calculate the accumulated value function $v_j(s_j)$.

*3.4. Lower level model*

Given the signal timings $\boldsymbol{\alpha}_j$ and $\boldsymbol{g}_j$ in barrier group $j$ from the middle level model, the lower level model determines the CAVs that cross the intersection during barrier group $j$, platoon these CAVs in the buffer zone, and plan their trajectories in the passing zone. The planned travel time $t_j^\omega$ is fed to the middle level model to calculate CAV delay.

3.4.1. CAV platooning

Given the signal timings $\boldsymbol{\alpha}_j$ and $\boldsymbol{g}_j$ in barrier group $j$, the start and end time points of each phase in barrier group $j$ are known. The CAVs that could cross the intersection during barrier group $j$ could be determined together with CAV platooning. **Fig. 5** shows the phase structure in a barrier group with north- and south-bound traffic as an example. The phase sequence is divided into two cases. The first is the head-lag left-turn phase sequence and the second includes the head-head and lag-lag left-turn phase sequences.

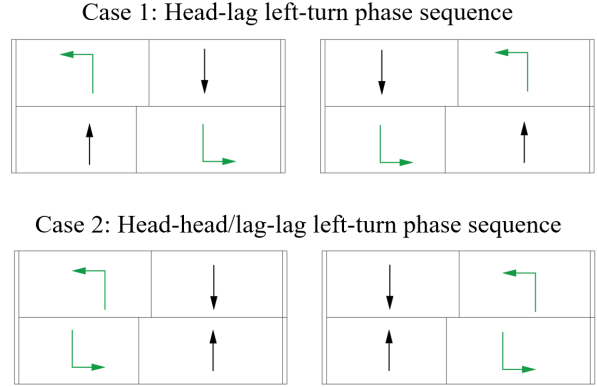

**Fig. 5** Two cases of phase sequences.

**Case 1: Head-lag left-turn phase sequence**

When the head-lag left-turn phase sequence is used, left-turn and through CAVs in each arm enter the passing zone according to the first-come-first-served rule. That is, the entering order is determined by their arrival times at the buffer zone. As shown in **Fig. 6**, CAVs that could cross the intersection in the same lane during the same barrier group are regarded as one platoon. The platoon-leading CAV of the platoon in barrier group $j$ in each arm is the first one of the CAVs that cannot cross the intersection in the previous $j - 1$ barrier groups. The trajectory planning for a platoon-leading CAV in the passing zone is described in Section 3.4.2, which makes the platoon pass the stop line without stops. The trajectories of platoon-following CAVs are captured by the car-following model in Section 3.4.3. As the signal timings in barrier group $j$ are fixed in the lower level model, the last platoon-following CAV in barrier group $j$ in each arm can then be determined. CAV platoons could wait in the buffer zone until the entry time into the passing zone. Therefore, the travel times $t_j^\omega$ of these CAVs in Case 1 are determined by Section 3.4.2 and Section 3.4.3. For the CAVs that cannot cross the intersection in barrier group $j$, the travel time is set as $t_j^\omega = t_{free}^\omega$.

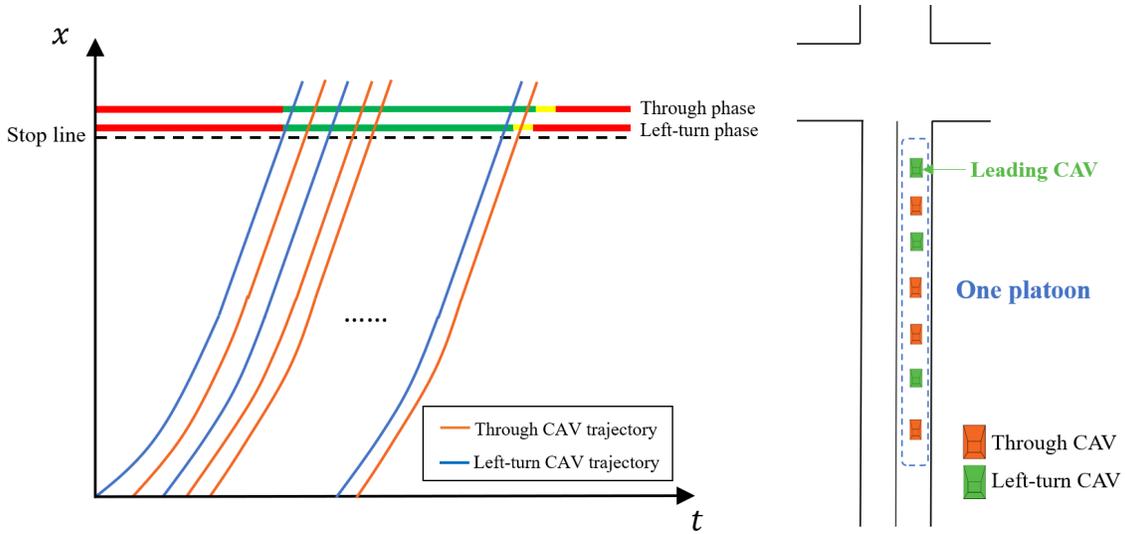

**Fig. 6** CAV trajectories and platooning in a barrier group in Case 1.

**Case 2: Head-head/lag-lag left-turn phase sequence**

When the head-head or lag-lag left-turn phase sequence is used, left-turn and through CAVs are reorganized into a left-turn CAV group and a through CAV group in the buffer zone. CAV platoons could further wait in the buffer zone until the entry time into the passing zone. For illustration, the lag-lag left-turn phase sequence is taken as an example, in which the left-turn CAV group follows the through CAV group to enter the passing zone in each arm. According to the signal timings, there are two subcases.

*Subcase 2-1: Two platoon.* As shown in **Fig. 7**(a), the first CAV of the left-turn group will come across the red light if it follows the last CAV of the through group. The left-turn group and the through group that could cross the intersection in one barrier group are then regarded as two platoons. The trajectories of the platoon-leading CAVs in the left-turn and through CAV platoons in the passing zone are planned in Section 3.4.2 to make the CAV platoons pass the stop line without stops. The trajectories of the following CAVs in the left-turn and through CAV platoons are captured by the car-following model in Section 3.4.3.

*Subcase 2-2: One platoons.* As shown in **Fig. 7(b)**, the first CAV of the left-turn group will pass the stop line without stops if it follows the last CAV of the through group. The left-turn group and the through group that could cross the intersection in one barrier group are then regarded as one platoon. The trajectory of the platoon-leading CAV in the passing zone is planned in Section 3.4.2 to make the CAV platoon pass the stop line without stops. The trajectories of the following through and left-turn CAVs in the platoon are captured by the car-following model in Section 3.4.3.

As the signal timings in barrier group $j$ are fixed in the lower level model, the last platoon-following CAVs of the through and left-turn CAV groups in barrier group $j$ in each arm can then be determined. Therefore, the travel times $t_j^\omega$ of these CAVs in Case 2 are determined by Section 3.4.2 and Section 3.4.3. For the CAVs that cannot cross the intersection in barrier group $j$, the travel time is set as $t_j^\omega = t_{free}^\omega$.

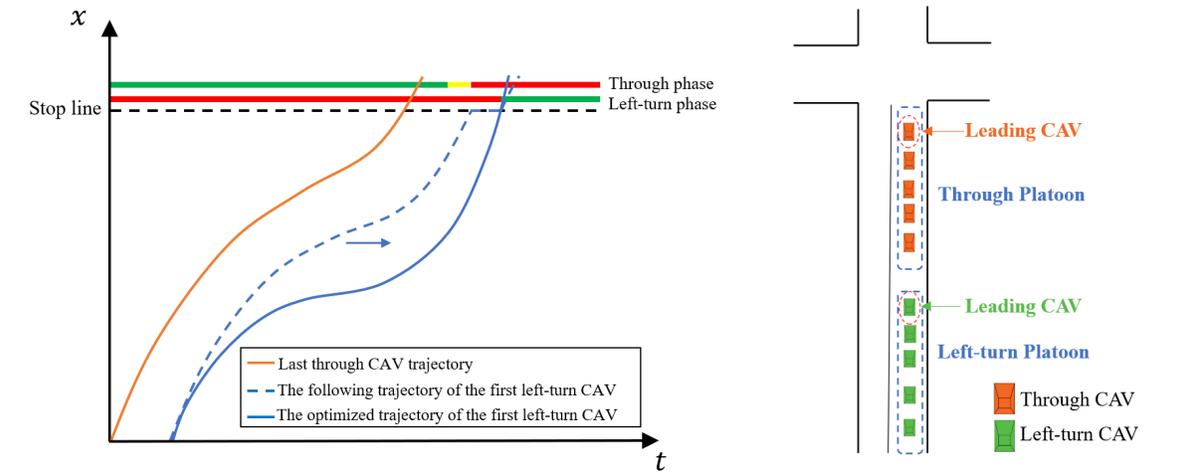

(a) Subcase 2-1 with two CAV platoons

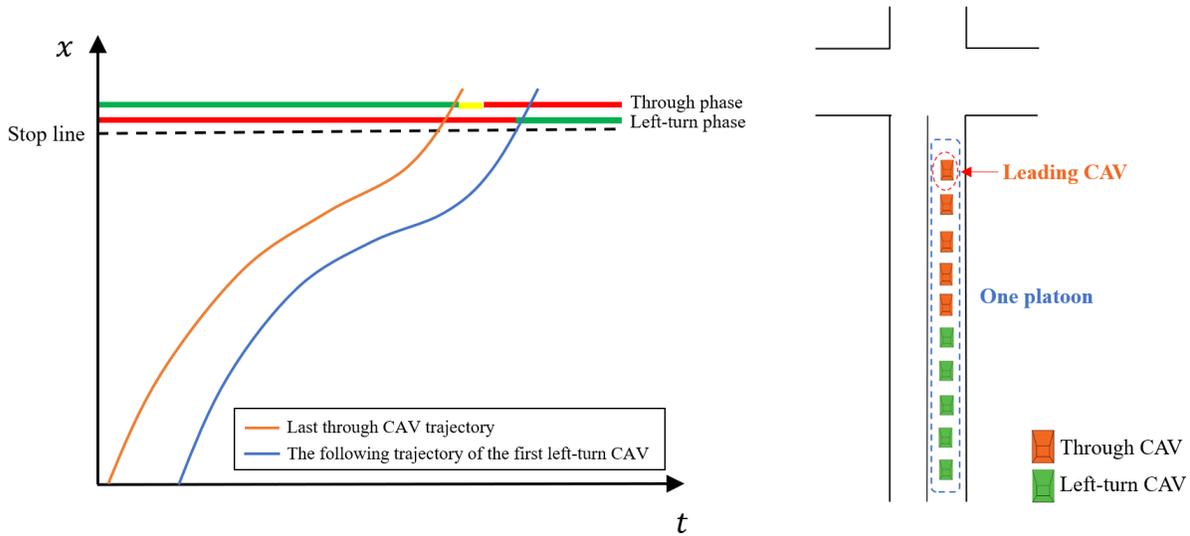

(b) Subcase 2-2 with one CAV platoon

**Fig. 7** CAV trajectories and platooning with lag-lag left-turn phase sequence.

### 3.4.2. Trajectory planning for platoon-leading CAVs

The acceleration profile of a platoon-leading CAV in the passing zone is optimized for minimum travel delay. Similar to the previous study (Feng et al., 2018), three-segment trajectory planning is applied for simplification as shown in **Fig. 8**. The trajectory is broken into three segments and a CAV travels with the maximum acceleration rate, maximum deceleration rate, or constant speeds on each trajectory segment. CAVs are guaranteed to pass the stop line without stops. The trajectory planning model (**P3**) is formulated based on continuous time:

(**P3**)

$$\min t_f^\omega \tag{25}$$

$$\max v^\omega(t_f^\omega) \tag{26}$$

s.t.

$$\begin{cases} x^\omega(t_0^\omega) = 0 \\ v^\omega(t_0^\omega) = v_0^\omega \end{cases} \tag{27}$$

$$x^\omega(t_f^\omega) = L \tag{28}$$

$$0 \leq v^\omega(t) \leq v_{max} \tag{29}$$

$$a_i^\omega \in \{a^U, 0, a^L\}, i = 1,2,3 \tag{30}$$

$$v^\omega(t_1^\omega) - v^\omega(t_0^\omega) = (t_1^\omega - t_0^\omega)a_1^\omega \tag{31}$$

$$v^\omega(t_2^\omega) - v^\omega(t_1^\omega) = (t_2^\omega - t_1^\omega)a_2^\omega \tag{32}$$

$$v^\omega(t_f^\omega) - v^\omega(t_2^\omega) = (t_f^\omega - t_2^\omega)a_3^\omega \tag{33}$$

$$x^\omega(t_1^\omega) - x^\omega(t_0^\omega) = \frac{(t_1^\omega - t_0^\omega)(v^\omega(t_1^\omega) + v^\omega(t_0^\omega))}{2} \tag{34}$$

$$x^\omega(t_2^\omega) - x^\omega(t_1^\omega) = \frac{(t_2^\omega - t_1^\omega)(v^\omega(t_2^\omega) + v^\omega(t_1^\omega))}{2} \tag{35}$$

$$x^\omega(t_f^\omega) - x^\omega(t_2^\omega) = \frac{(t_f^\omega - t_2^\omega)(v^\omega(t_f^\omega) + v^\omega(t_2^\omega))}{2} \tag{36}$$

$$t_{p,r,j}^g + g_{p,r,j}\Delta t \geq t_f^\omega \geq t_{p,r,j}^g \tag{37}$$

where $t_0^\omega$ is the entry time of CAV $\omega$ into the passing zone; $t_1^\omega$ and $t_2^\omega$ are the end times of the 1st and 2nd trajectory segments; $t_f^\omega$ is the time point when CAV $\omega$ passes the stop line; $v^\omega(t)$ and $x^\omega(t)$ are the speed and travelled distance of CAV $\omega$ at time point $t$, respectively; $a_i^\omega$ is the acceleration/deceleration rate of CAV $\omega$ at the $i^{th}$ trajectory segment; $a^U/a^L$ is the maximum acceleration/deceleration rate; $v_0^\omega$ is the entry speed of CAV $\omega$ into the passing zone; $L$ is the length of the passing zone; $t_{p,r,j}^g$ is the start time point of phase $p$ in ring $r$ in barrier $j$ in which CAV $\omega$ crosses the intersection. The decision variables are $t_i^\omega$ ($i = 1,2$), $t_f^\omega$, $a_i^\omega$ ($i = 1,2,3$), $v^\omega(t)$, and $x^\omega(t)$. The others are known parameters.

Since the free-flow travel time is a constant, minimizing travel delay equals minimizing the planned travel time, which is the difference between $t_f^\omega$ and the arrival time. The arrival time is a constant and therefore minimization of $t_f^\omega$ is adopted as the primary objective function Eq. (25). It is noted that multiple optimal solutions may exist with the same passing time $t_f^\omega$ as shown in **Fig. 9.** The trajectories with higher speeds of passing the stop line are preferred because higher passing speeds indicate less travel time within the intersection area and exit lanes. The maximization of the passing speed $v^\omega(t_f^\omega)$ is used as the secondary objective Eq. (26). The entry time $t_0^\omega$ is related to the applied rolling horizon scheme and is discussion in Section 3.5.2. The initial and final states at $t_0^\omega$ and $t_f^\omega$ are defined in Eqs. (27) and (28). Eq. (29) set the lower and upper bounds of the speed. Eq. (30) indicates the acceleration rate takes $a^U$, 0, or $a^L$. Eqs. (31)-(33) build the relationship between the speeds and accelerate rate of the three trajectory segments. Eqs. (34)-(36) build the relationship between the speeds and travelled distance of the three trajectory segments. Eq. (37) guarantees that CAV $\omega$ crosses the intersection during the green time of the corresponding phase. Note that the entry speed $v_0^\omega$ is unknown before the entry time $t_0^\omega$. It is assumed that $v_0^\omega$ has a lower bound $v_0^L$ and an upper bound $v_0^U$. In the signal timings, $v_0^U$ is taken to replace $v_0^\omega$ for the trajectory planning (i.e., $v_0^\omega = v_0^U$) because vehicles prefer higher speeds. When CAV platoons enter the passing zone at $t_0^\omega$, the trajectories will be re-planned based on the actual entry speed. The proposed trajectory planning model (**P3**) is a nonlinear programming model with a hierarchical multi-objective optimization structure. It can be solved efficiently by the solution algorithm in Appendix A. Note that the optimal trajectory may have less than three segments.

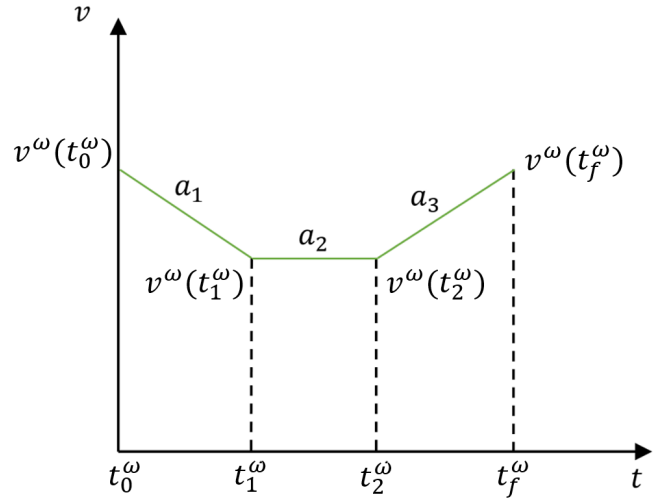

**Fig. 8** Three-segment trajectory planning for the platoon-leading CAV.

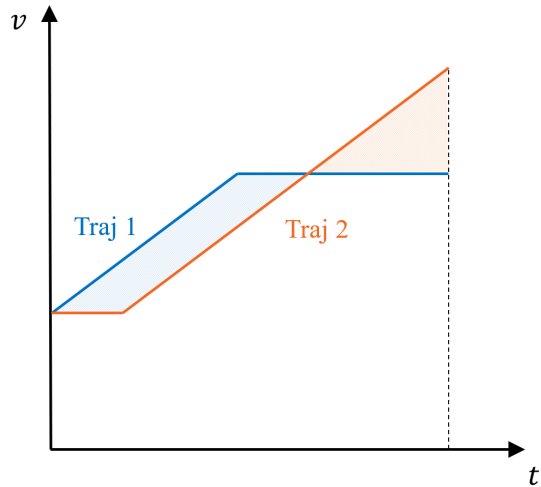

**Fig. 9** Two exemplary trajectories with the same travel time.

3.4.3. Car-following model for platoon-following CAVs

The trajectories of platoon-following CAVs are captured by the Next Generation Simulation (NGSIM) car-following model (Yeo et al., 2008; Feng et al., 2018). This model takes into consideration vehicle performance limits such as the maximum acceleration and deceleration rates. It is formulated as

$$x^\omega(t+\Delta t) = \max\{x_\omega^U(t+\Delta t), x_\omega^L(t+\Delta t)\} \tag{38}$$

where $x_\omega^L(t)$ and $x_\omega^U(t)$ are the lower and upper bounds of $x^\omega(t)$, respectively.

The upper bound is related to the vehicle-following rules, maximum acceleration/deceleration rates, speed limits, and maximum safety distance for collision avoidance. $x_\omega^U(t + \Delta t)$ can be expressed as

$$x_\omega^U(t + \Delta t) = \min\{x^{\omega'}(t + \Delta t - \tau^\omega) - l^{\omega'} - d_{jam}^\omega, x^\omega(t) + v^\omega(t)\Delta t + a^U \Delta t^2, x^\omega(t) + v_{max}\Delta t,\ x^\omega(t) + \Delta x^\omega(t + \Delta t)\} \tag{39}$$

$$\Delta x^\omega(t + \Delta t) = \Delta t \left( a^L \tau^\omega + \sqrt{(a^L \tau^\omega)^2 - 2a^L \left( x^{\omega'}(t) - x^\omega(t) - l^{\omega'} - d_{jam}^\omega - \frac{(x^{\omega'}(t))^2}{2a^L} \right)} \right) \tag{40}$$

where $\omega'$ is the leading vehicle of CAV $\omega$; $\tau^\omega$ is the reaction time of CAV $\omega$; $d_{jam}^\omega$ is the jam spacing between CAVs $\omega$ and $\omega'$; $l^{\omega'}$ is the length of CAV $\omega'$; $\Delta x^\omega(t + \Delta t)$ is the maximum safety distance that must be maintained to avoid collision of CAV $\omega$ at time $t + \Delta t$.

The lower bound is related to the maximum acceleration/deceleration rates and the current location to prevent the CAV from going backwards. $x_\omega^L(t + \Delta t)$ is expressed as

$$x_\omega^L(t + \Delta t) = \max\{x^\omega(t), x^\omega(t) + v^\omega(t)\Delta t + a^L \Delta t^2\} \tag{41}$$

*3.5. Discussion on the buffer zone and the passing zone*

3.5.1. Length of the passing zone

CAVs are platooned in the buffer zone and could wait for the entry time into the passing zone. It is assumed that CAV platoons enter the passing zone from the buffer zone at a speed within the range $[v_0^L, v_0^U]$. The length of the passing zone should be set properly to make the trajectory planning model feasible in Section 3.4.2 and guarantee CAV platoons to cross the intersection without stops at the stop line. Please refer to the previous studies (Feng et al., 2018; Yu et al., 2018) for the analytical analysis of the impacts of travel distance on CAV trajectory planning. It is noted that CAV platooning (including lane-changing maneuvers) in the buffer zone may affect the arrival pattern of HVs at the stop line, especially when the buffer zone is close to the stop line. This influence makes the estimation of HV delay inaccurate in Section 3.3.1. However, a too long length of the passing zone makes it difficult to coordinate CAV trajectory planning and signal timing, which increases model complexity. Considering these factors, the recommended length $L_s$ of the passing zone is defined in Eq. (42). The definition of $L_s$ is related to the applied rolling horizon scheme in Section 3.5.2.

$$L_s = \frac{v_{max}^2 - v_0^{L^2}}{2a^U} + (X_j^{min} - \frac{v_{max} - v_0^L}{a^U})v_{max} \tag{42}$$

3.5.2. Elimination of the buffer zone

The setting of the buffer zone may restrict the application of the proposed model. It is noted that the buffer zone could be eliminated when the head-lag left-turn phase sequence is used and the through and left-turn phases for traffic from the same arm start and end at the same time as shown in **Fig. 10**. In that case, the reorganization of left-turn CAV platoons and through CAV platoons is not necessary. Left-turn and through CAVs could cross the intersection in one mixed platoon as shown in **Fig. 6**. The order of CAVs crossing the intersection is determined by their arrival times, i.e., the first-come-first-served rule. Consequently, CAV platoons do not need to wait in the buffer zone. Additionally, the trajectory planning for platoon-leading CAVs could be conducted without the buffer zone.

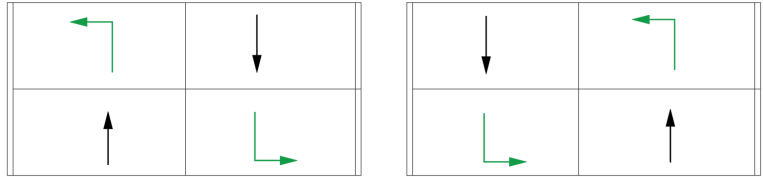

**Fig. 10** Head-lag left-turn phase sequence with synchronized through and left-turn phases.

## 4. Rolling horizon scheme

A rolling horizon scheme is designed to apply the proposed model dynamically under time-varying traffic conditions as shown in **Fig. 11**. The signal timings of the first two barrier groups (i.e., the first signal cycle) are applied each time the proposed model is solved based on updated traffic conditions. During the first barrier group, a new optimization procedure is conducted with newly predicted/collected arrival information of HVs and CAVs. The first signal cycle of the new optimization is then applied after the current signal cycle. Take the scenario in **Fig. 11** as an example. During the first barrier group (i.e., east- and west-bound traffic phases) of the current cycle (i.e., cycle one), a new optimization is conducted. The new optimization determines the signal timings of cycle two and the trajectory planning for CAVs that could cross the intersection in cycle two. At the end time $t_{EW}$ of the first barrier group of cycle one, CAVs in the east- and west-bound arms leave the buffer zone and enter the passing zone. CAV trajectories are re-planned based on the actual entry speed and the signal timings. CAVs will follow the re-planned trajectories to cross the intersection during the first barrier group of cycle two. At the end time $t_{NS}$ of the second barrier group (i.e., north- and south-bound traffic phases) of cycle one, CAVs in the north- and south-bound arms leave the buffer zone and enter the passing zone with re-planned trajectories. CAVs will cross the intersection during the second barrier group of cycle two. During the first barrier group of cycle two, a new optimization is then conducted and similar procedures are repeated.

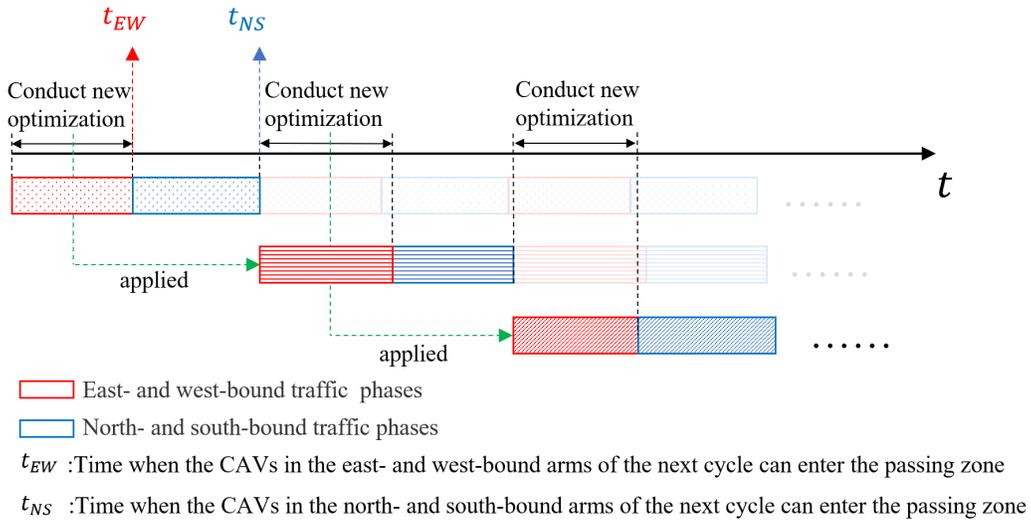

**Fig. 11** Rolling horizon scheme and entry time of CAV platoons into the passing zone.

## 5. Numerical studies

*5.1. Simulation setup*

To evaluate the performance of the proposed model, a typical four-arm intersection with all directions of movements is applied. **Fig. 12** shows the lane markings of the intersection. Right-turn vehicles are not controlled by traffic signals. The length of the passing zone is $L = 500$ m. It is assumed that the communication range is sufficient for the intersection traffic control. For example, the target communication range of DSRC could reach 1 km (Kenney, 2011).

In the upper level model, minimum green time $G_{p,r,j}^{min} = 15$ s, maximum green time $G_{p,r,j}^{max} = 25$ s, and green interval $R_{p,r,j} = 4$ s. In the middle level model, the saturation flow rate of HVs $q_m^s$ are 1550 veh/h and 1650 veh/h when $m = 2, 4, 5, 8$ (i.e., left turn) and $m = 1, 3, 6, 7$ (i.e., through). In the lower level model, the maximum acceleration rates $a^U = 2$ m/s², maximum deceleration rates $a^L = -2$ m/s², maximum speed $v_{max} = 14$ m/s, sum of jam spacing and vehicle $l_\omega + d_\omega^{jam} = 6$ m, and reaction time $\tau_\omega = 0$ s.. The initial speeds of platoon-leading CAVs entering the passing zone are randomly generated between $v_0^L = 3$ m/s and $v_0^U = 14$ m/s. VISSIM 4.3 (PTV AG, 2008) is used for the simulation. The Wiedemann 74 car-following model and the Free Lane Selection model in VISSIM 4.3 are used to capture the driving behaviors of HVs in the simulation. The same maximum acceleration/deceleration rate and maximum speed are set for HVs and CAVs. The reaction time of HVs is 0.5 s.

The average vehicle arrival rate is 2250 veh/h in each arm. The uniform and Poisson distributions of vehicle arrivals are both tested. The proportions of left-turn, through, and right-turn traffic are 0.4, 0.4, and 0.2, respectively. The CAV penetration rate is 50%. Left-turn and through CAVs share the CAV-dedicated lanes to cross the intersection and right-turn CAVs share the right-most lanes with HVs. The prediction horizon $T = 120$ s. For simplification, the average arrival rate of HVs in the prediction horizon is predicted as the one measured in the last $T$ seconds. The arrival time of each CAV in the prediction horizon could be collected by communication technologies.

The BP-based control model in Rey and Levin (2019) is applied as the benchmark as shown in **Fig. 13**, in which CAV-dedicated phases and lanes are used. Each barrier group is followed by a CAV-dedicated phase, whose duration is set to the minimum green time. There is no CAV trajectory planning in approaching lanes and point-queues are used. The implemented models are written in Python 3.7.4. All the experiments are performed in a desktop computer with an Intel 2.6 GHz CPU and 16 GB memory. The parallel computing with five threads is used for computational efficiency. The average computational time of solving the proposed model is ~2 s. Five random seeds are used in the simulation for each experiment considering the uncertainty in vehicle arrivals and HV driving behaviors. Each simulation run is 1000 s with a warm-up period of 120 s. The length of a time step $\Delta t = 1$ s.

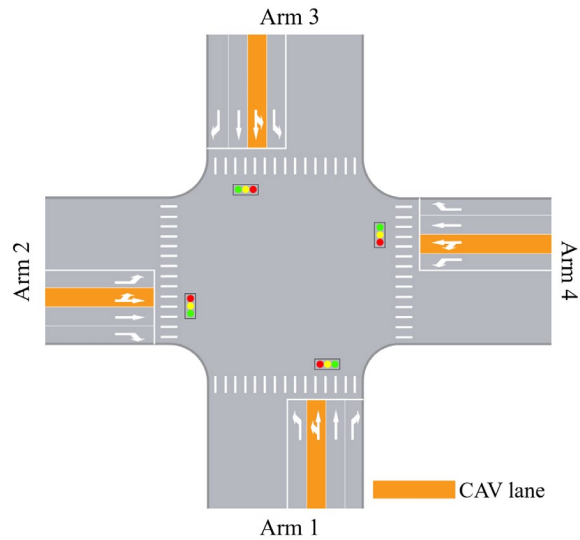

**Fig. 12.** Lane markings of a four-arm intersection.

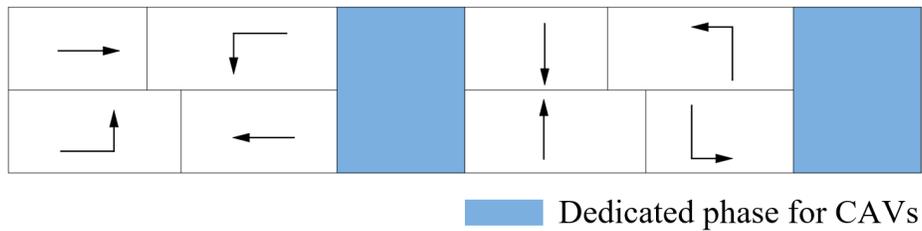

**Fig. 13** BP-based control.

5.2. *Simulation results and discussion*

Average vehicle throughput and vehicle delay are used as the performance measures to compare the performance of the proposed SPDL-based control model and the benchmark BP-based control model. The delay of a vehicle is calculated as the difference between the actual travel time and the free-flow travel time. The travel time is the time of leaving the intersection minus the time of arriving at the intersection. Only the delays of the vehicles that have left the intersection are counted. The simulation results are shown in **Fig. 14**-**Fig. 15**.

**Fig. 14** shows the vehicle throughput under the SPDL-based control and the BP-based control. The throughput of all vehicles, HVs, and CAVs are illustrated in **Fig. 14**(a) and (b) with uniform and Poisson arrivals. **Fig. 14**(a) shows that the SPDL-based control improves the throughput of all vehicles, HVs, and CAVs by 2%, 6%, and 1%, respectively, compared with the BP-based control when vehicle arrivals conform to the uniform distribution. **Fig. 14**(b) shows that these throughput increases are 3%, 5%, and 2% when vehicle arrivals conform to the Poisson distribution. The avoidance of CAV-dedicated phases helps reduce the clearance lost time. For example, as shown in **Fig. 16**, there is more all-red clearance time in one cycle under the BP-based control compared to that under the SPDL-based control. Further, CAV-dedicated phases could be regarded as a kind of lost time to HVs. As a result, the HV capacity is improved by the SPDL-based control with shared phases. Such benefits are more noticeable with higher demand as shown in Section 5.3.1. The introduction of trajectory planning for CAVs make CAVs cross the intersection without stops at the stop line as shown in **Fig. 17**. In this way, the start-up lost time is eliminated under the SPDL-based control. The CAV capacity is thus improved by the SPDL-based control but the improvement is slight. Because the start-up lost time (e.g., 2 s) is usually a small value compared to the cycle length.

**Fig. 15** shows the average vehicle delay under the SPDL-based control and the BP-based control. The average delays of all vehicles, HVs, and CAVs are illustrated in **Fig. 15**(a) and (b) with uniform and Poisson arrivals. **Fig. 15**(a) shows that the SPDL-based control reduces the delays of all vehicles, CAVs, and HVs by 47%, 1%, and 59%, respectively, compared with the BP-based control when vehicle arrivals conform to the uniform distribution. **Fig. 15**(b) shows that these delay reductions are 51%, 8%, and 62% when vehicle arrivals conform to the Poisson distribution. One noticeable observation is the significantly reduced delay of HVs (~60%). The main reason is that sharing phases with CAVs helps HVs reduce time-in-queue delay. For illustration, the results of signal timings in one cycle are taken as an example, as shown in **Fig. 16**. During CAV-dedicated phases, all HVs have to queue, which increase HV delay significantly. In contrast, the delay reduction of CAVs is less than 10%. The eliminated start-up lost time by the trajectory planning for CAVs could reduce CAV delay, but not as large as the reduction of HV delay. Note that the BP-based control in Rey and Levin (2019) uses point-queues, which means the advantages of the SPDL-based control is underestimated in terms of CAV delay.

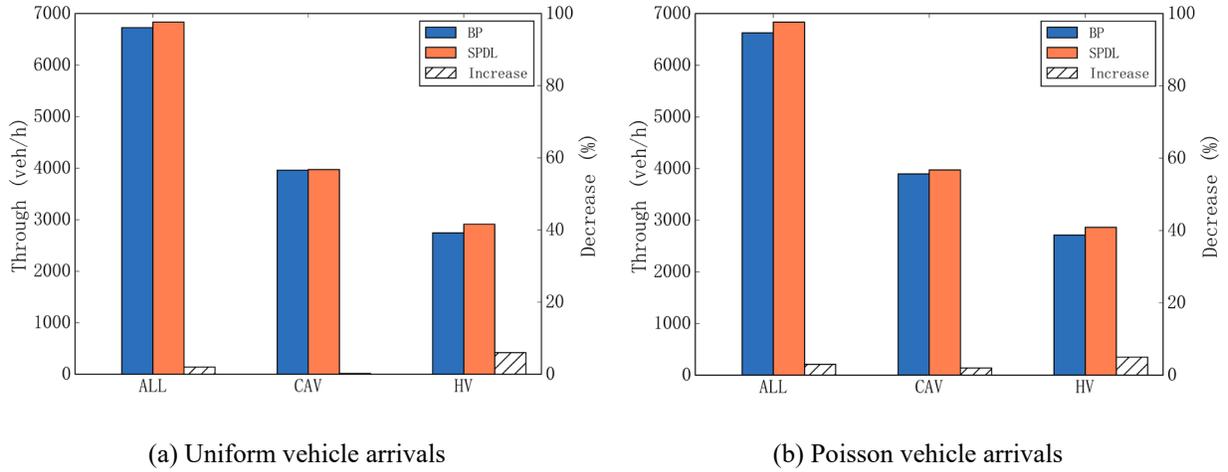

(a) Uniform vehicle arrivals  (b) Poisson vehicle arrivals

**Fig. 14** Throughput (veh/h).

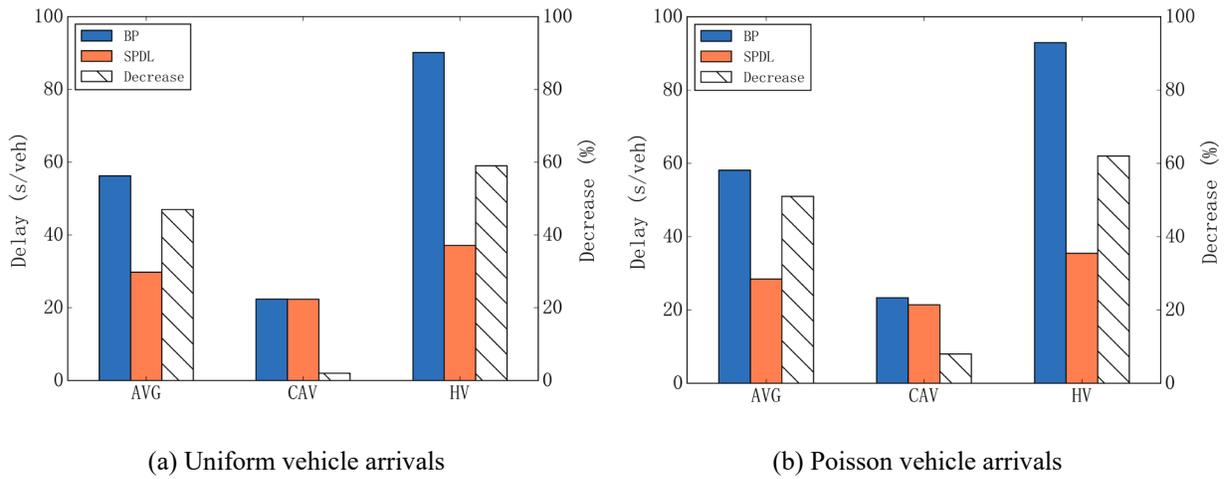

(a) Uniform vehicle arrivals  (b) Poisson vehicle arrivals

**Fig. 15** Average vehicle delay (s/veh).

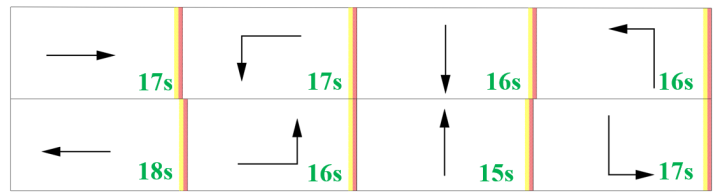

(a) Signal timings in one cycle under the SPDL-based control (with shown green durations)

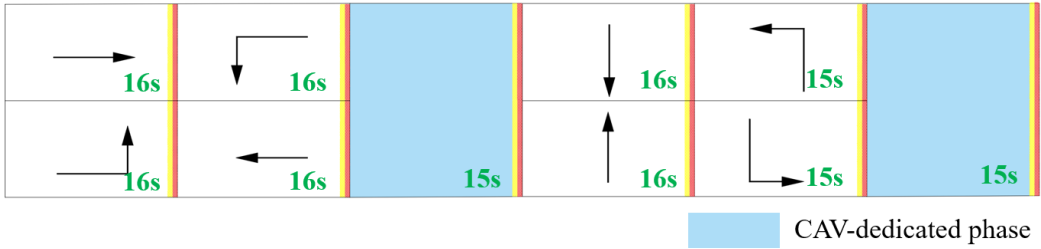

(b) Signal timings in one cycle under the BP-based control (with shown green durations)

**Fig. 16** Signal timings in one cycle.

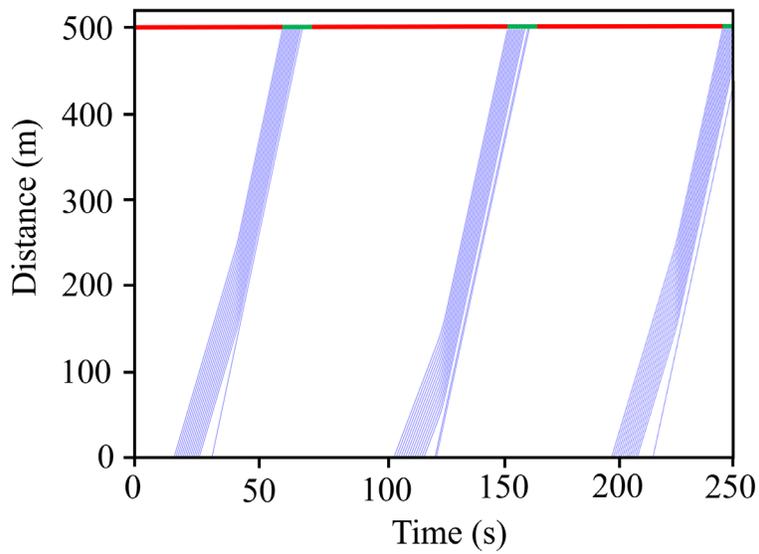

**Fig. 17** CAV trajectories under the SPDL-based control.

*5.3. Sensitivity analysis*

5.3.1. Traffic demand

Nine traffic demand levels are tested. The input demand is the product of a demand factor and the basic traffic demand in **Table 2**. The demand factor ranges from 0.5 to 4.5 at the step of 0.5. The traffic is undersaturated when the demand factor is less than or equal to 1.5. The traffic is oversaturated when the demand factor is more than or equal to 2.0. The undersaturated and oversaturated traffic conditions are observed under the BP-based control. The CAV penetration rate is set to 50%.

**Fig. 18** shows the vehicle throughput with increasing demand under the SPDL-based control and the BP-based control. The results with uniform and Poisson vehicle arrivals are shown in **Fig. 18**(a) and **Fig. 18**(b), respectively. When the traffic demand is undersaturated (with the demand factor $\leq$ 1.5), the vehicle throughputs under the SPDL-based control and the BP-based control are nearly the same as the demand. This indicates the intersection capacity is not reached and all demand can be accommodated. Oversaturated traffic can be observed when the throughput fails to catch the demand. Under the BP-based control, the throughput keeps increasing but deviates from the demand line when the demand factor increases from 1.5 to 3.0. That means, the traffic of partial phases is oversaturated. When the demand factor further increases from 3.0, the throughput line becomes flat. Because the intersection capacity is reached and no more demand could be accommodated. In contrast, the very demand factor is 3.5 under the SPDL-based control. The intersection capacity under the SPDL-based control is improved by ~10% compared to that under the BP-based control. As discussed in Section 5.2, there are two main reasons: 1) the avoidance of CAV-dedicated phases under the SPDL-based control could reduce the clearance lost time; and 2) the trajectory planning for CAVs could reduce the start-up lost time.

**Table 2** Basic traffic demand.

| | Traffic demand in pcu/h | | | | |
|---|---|---|---|---|---|
| | Left-turn | | Through | | Right-turn |
| | CAV | HV | CAV | HV | |
| Arm 1 | 200 | 200 | 200 | 200 | 100 |
| Arm 2 | 200 | 200 | 200 | 200 | 100 |
| Arm 3 | 200 | 200 | 200 | 200 | 100 |
| Arm 4 | 200 | 200 | 200 | 200 | 100 |

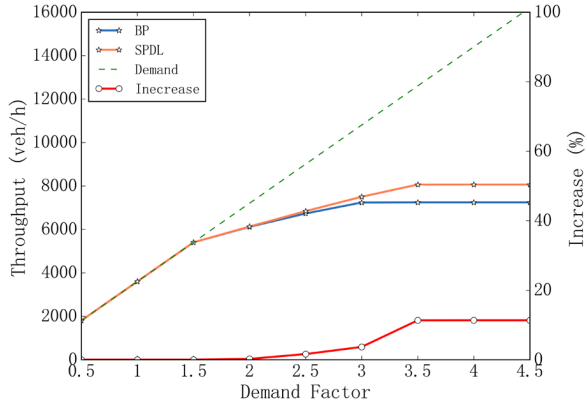
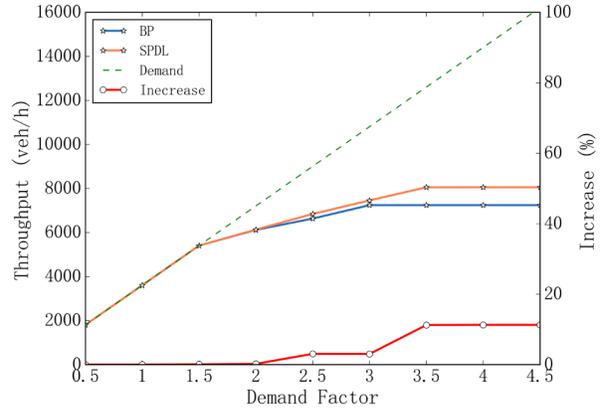

(a) Uniform vehicle arrivals  (b) Poisson vehicle arrivals

**Fig. 18** Vehicle throughput (veh/h).

**Fig. 19** shows the average vehicle delay with increasing demand under the SPDL-based control and the BP-based control. The results with uniform and Poisson vehicle arrivals are shown in **Fig. 19**(a) and **Fig. 19**(b), respectively. When the traffic demand is undersaturated (with the demand factor ≤ 1.5), the average vehicle delay increases slightly with increasing demand under both the SPDL- and BP-based control. The delay under the SPDL-based control is reduced by ~45% compared to that under the BP-based control. When the demand further increases, the average vehicle delay rises remarkably under both the control methods. For example, when the demand factor increases from 1.5 to 3.5, the delay increases under the BP- and SPDL-based control are 99% and 78%, respectively. However, the delay under the SPDL-based control approximately keeps half of that under the BP-based control at different demand levels. **Fig. 20** further shows the average vehicle delay of CAVs and HVs with increasing demand. One noticeable observation is the significant delay reduction of HVs under the SPDL-based control, especially with oversaturated traffic. For example, the delay reduction of HVs is ~60% when the demand factor is 3.5. The benefits are due to the avoidance of CAV-dedicated phases, which reduces the time-in-queue delay of HVs. In contrast, the CAV delays under the SPDL- and BP-based control differ insignificantly. When the traffic is not heavily saturated (with the demand factor < 3), the SPDL-based control slightly outperforms the BP-based control in terms of the CAV delay. But when the heavily saturated traffic is imposed (with the demand factor ≥ 3), the BP-based control slightly outperforms the SPDL-based control. These observations are consistent with our intuition about the advantages and disadvantages of CAV-dedicated phases.

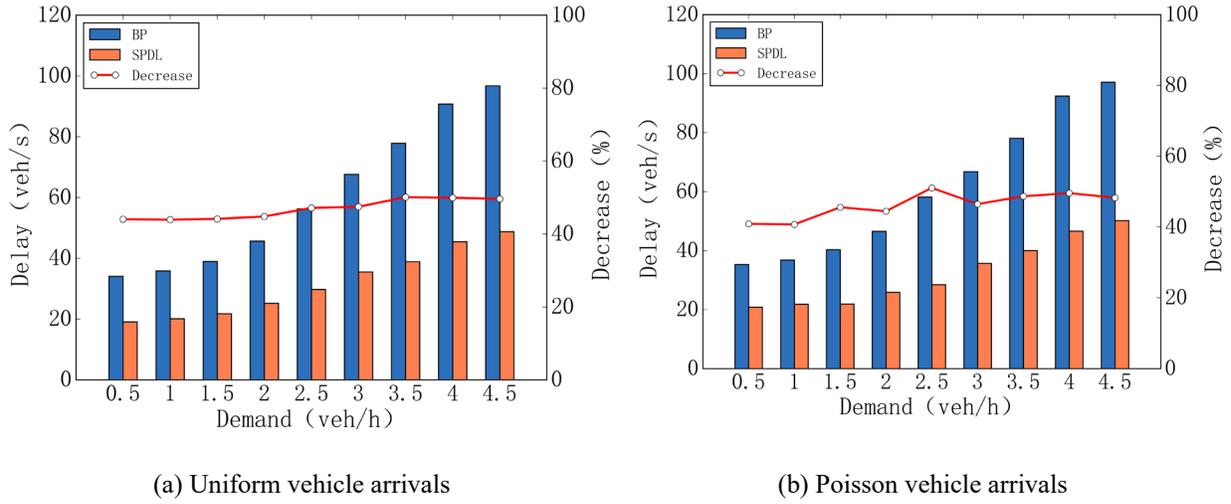

(a) Uniform vehicle arrivals    (b) Poisson vehicle arrivals

**Fig. 19** Average vehicle delay of all vehicles.

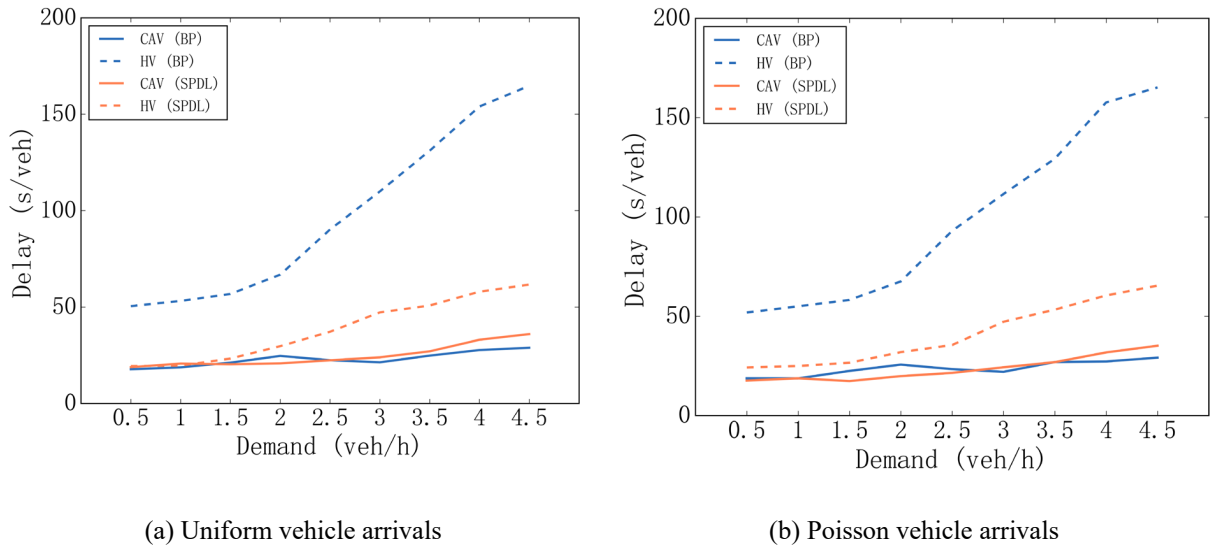

(a) Uniform vehicle arrivals    (b) Poisson vehicle arrivals

**Fig. 20** Average vehicle delay of CAVs and HVs.

5.3.2. CAV penetration rates

The tested CAV penetration rates range from 0% to 100% at the step of 10%. Four traffic demand levels are used, namely, 3600 veh/h, 7200 veh/h, 9000 veh/h, and 13500 veh/h. The corresponding demand factors are 1, 2, 2.5, and 3.75, which indicate uncongested traffic, slightly congested traffic, moderately congested traffic, and heavily congested traffic under the BP-based control, respectively. Both uniform and Poisson vehicle arrivals are applied.

**Fig. 21** shows that average vehicle delays with the increasing CAV penetration rate at the four demand levels when the uniform distribution of vehicle arrivals is used. Generally, a high penetration rate of CAVs is beneficial to the delay reduction under both the BP- and SPDL-based control. When the CAV penetration rate is less than 70%, the average delay is large and decreases remarkably with the increasing CAV penetration rate under the BP-based control, especially when the traffic is heavily congested. For example, as shown in **Fig. 21**(d), the average delay decreases from ~130 s/veh to ~45 s/veh by ~65% under the BP-based control when the penetration rate increases from 10% to 70%. However, the delay decrease becomes insignificant at all the demand levels when the penetration rate further increases. This indicates a high CAV penetration rate ≥ 70% is required to guarantee the performance of the BP-based control. In contrast, the SPDL-based control achieves much less delay with both low and high CAV penetration rates. And the benefits are more noticeable with low penetration rates. For example, the delay reduction reaches ~60% with heavily congested traffic when the penetration rate is 10% as shown in **Fig. 21**(d). In addition, the performance of the SPDL-based control is relatively more robust to the penetration rate than the BP-based control at all the demand levels. Similar observations can be obtained in **Fig. 22** with Poisson vehicle arrivals.

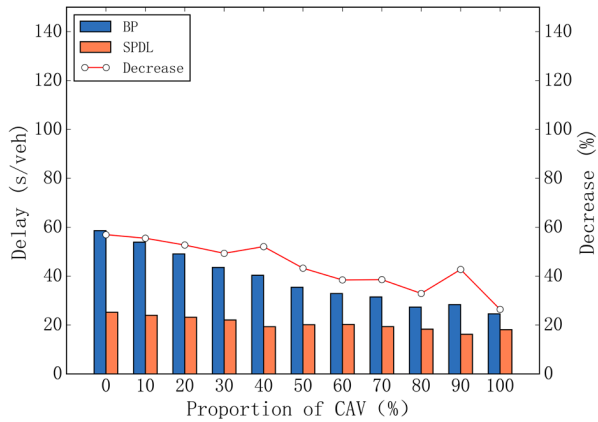

(a) Demand factors=1

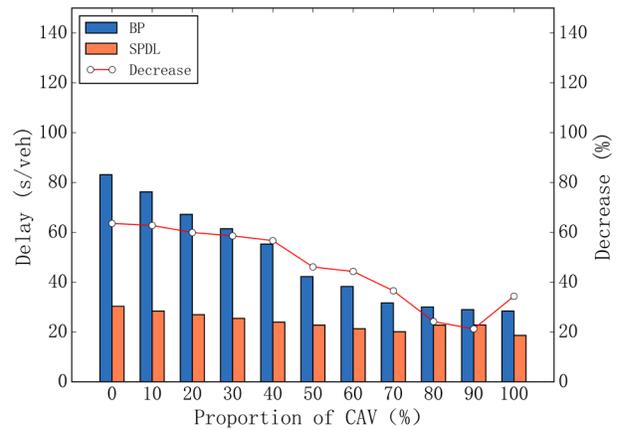

(b) Demand factors=2

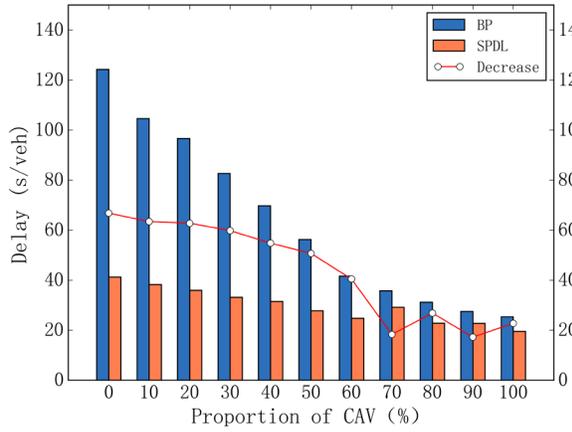

(c) Demand factors=2.5

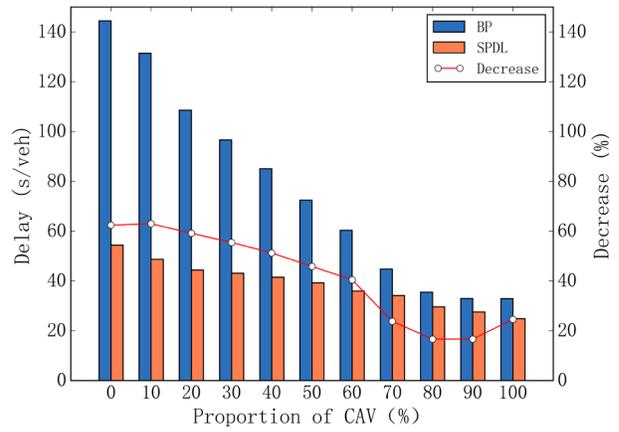

(d) Demand factors=3.75

**Fig. 21** Impacts of CAV penetration rates on vehicle delay with uniform vehicle arrivals.

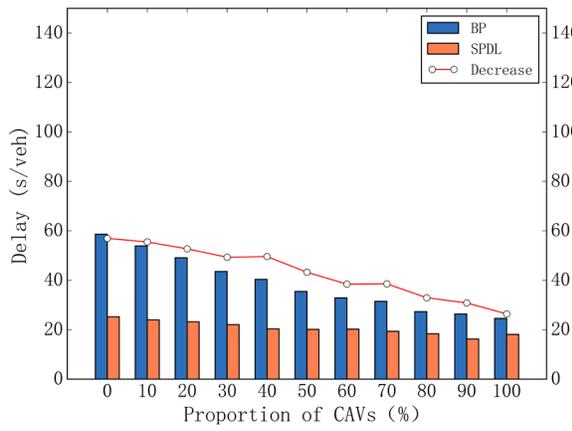

(a) Demand factors=1

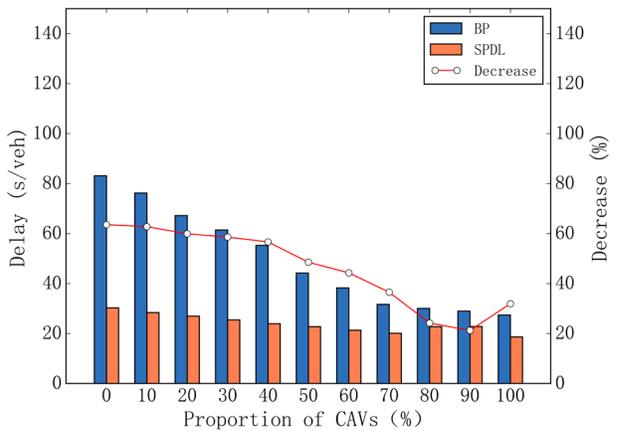

(b) Demand factors=2

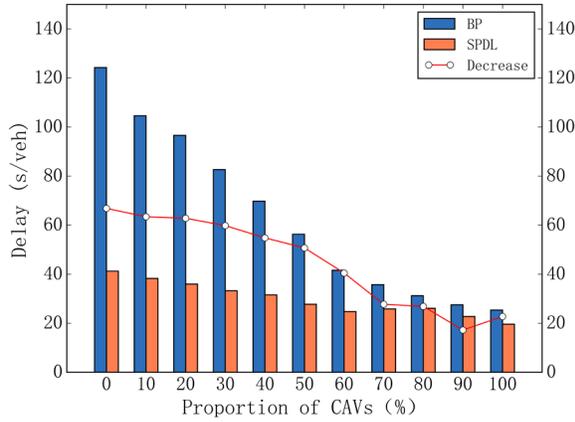 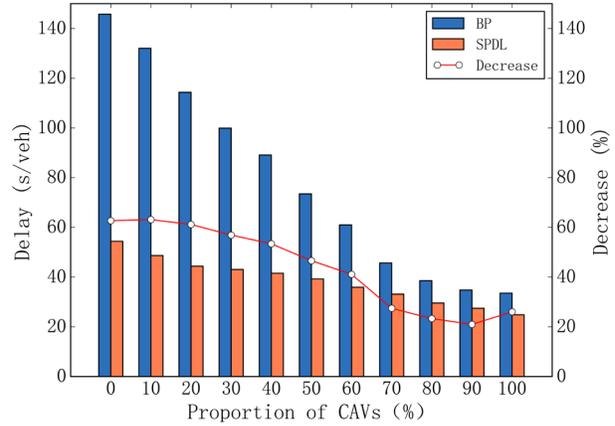

(c) Demand factors=2.5  (d) Demand factors=3.75

**Fig. 22** Impacts of CAV penetration rates on vehicle delay with Poisson vehicle arrivals.

5.3.3. Left-turn proportions

The tested left-turn proportions range from 0% to 100% at the step of 10%. Four traffic demand levels are used, namely, 3600 veh/h, 7200 veh/h, 9000 veh/h, and 12600 veh/h. The corresponding demand factors are 1, 2, 2.5, and 3.5. Vehicle arrivals conform to the Poisson distribution. The CAV penetration rate is 50%. Only through and left-turn vehicles are considered. **Fig. 23** shows the average vehicle delays with the increasing left-turn proportion at the four demand levels. When the left-turn proportion is less than 50%, the average delay decreases with the increasing left-turn proportion under both the SPDL- and BP-based control. When the left-turn proportion is more than 50%, the average delay increases with the increasing left-turn proportion under both control. That is, the average delay reaches the smallest at the left-turn proportion of 50%. In that case, traffic demand in through and left-turn HV lanes is balanced. Moreover, it is observed that the delay reduction by the SPDL-based control is the most significant compared with the BP-based control at the left-turn proportion of 50%. In that case, through and left-turn CAVs from the four arms have the most severe conflict within the intersection area during the CAV-dedicated phases under the BP-based control. In contrast, the CAV conflict is alleviated by the phase sharing with HVs and CAVs under the SPDL-based control. Because only CAVs with compatible movements travel within the intersection area at the same time.

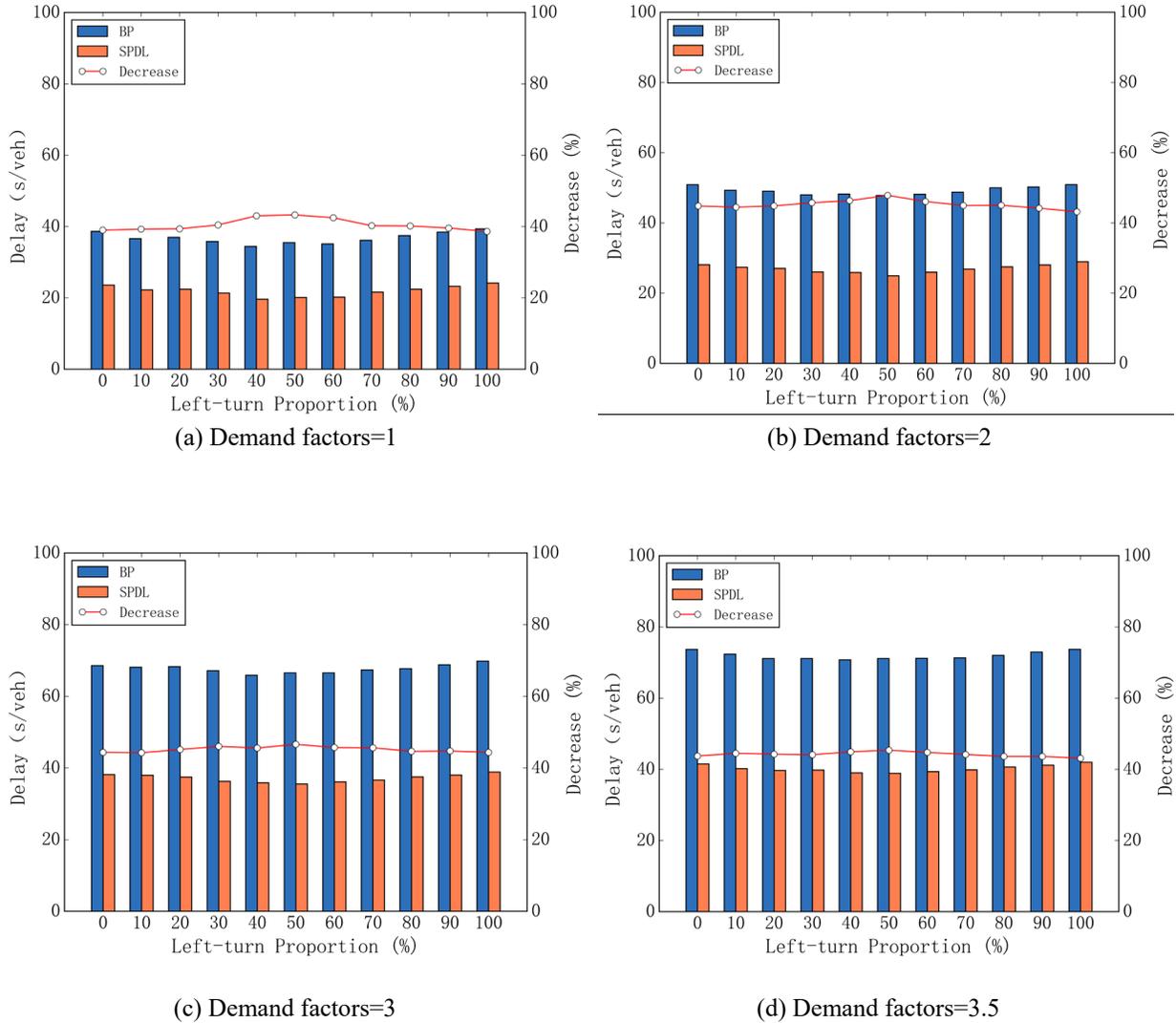

(a) Demand factors=1  (b) Demand factors=2

(c) Demand factors=3  (d) Demand factors=3.5

**Fig. 23** Impacts of left-turn proportions on vehicle delay with Poisson vehicle arrivals.

### 5.4. Extended model without the buffer zone

As discussed in Section 3.5.2, the buffer zone could be removed if the head-lag left-turn phase sequence in **Fig. 10** is used. This section explores the performance of the extended SPDL-based control model without the buffer zone. The four demand levels in Section 5.3.2 are used (3600 veh/h, 7200 veh/h, 9000 veh/h, and 13500 veh/h), which cover uncongested traffic, slightly congested traffic, moderately congested traffic, and heavily congested traffic. A low CAV penetration rate of 20% and a high CAV penetration rate of 80% are tested. Other parameters are the same as those in Section 5.1.

**Fig. 24** and **Fig. 25** show the average vehicle delay under the SPDL-based control, the extended SPDL-based control, and the BP-based control with the low and high CAV penetration rates, respectively. **Fig. 24** shows that the extended SPDL-based control significantly outperforms the BP-based control at all the demand levels when the CAV penetration rate is 20%. The delay reduction reaches ~50% with the uncongested traffic and the benefits are more remarkable with higher demand. Since the phase structure of the extended SPDL-based control is fixed, its performance is worse than that of the SPDL-based control. The extended SPDL-based control raises the delay by less than 10% compared to the SPDL-based control. But it still significantly outperforms the BP-based control. **Fig. 25** shows the advantages of the extended SPDL-based control over the BP-based control with the high CAV penetration rate of 80%, which are not as large as those in **Fig. 24** with the low penetration rate. There observations are consistent with the ones in Section 5.3.2. Compared with the SPDL-based control, the extended SPDL-based control raises vehicle delay by ~10% in **Fig. 25**. All in all, the extended SPDL-based control could serve as an alternative approach when it is difficult to set the buffer zone.

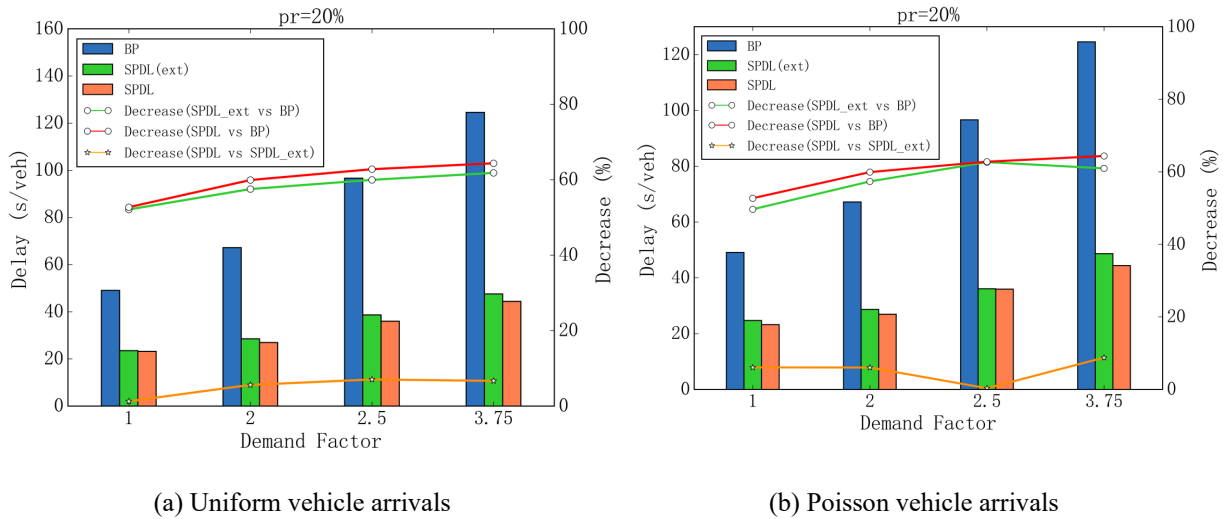

(a) Uniform vehicle arrivals     (b) Poisson vehicle arrivals

**Fig. 24.** Average vehicle delays of the extended model with the CAV proportion rate of 20%.

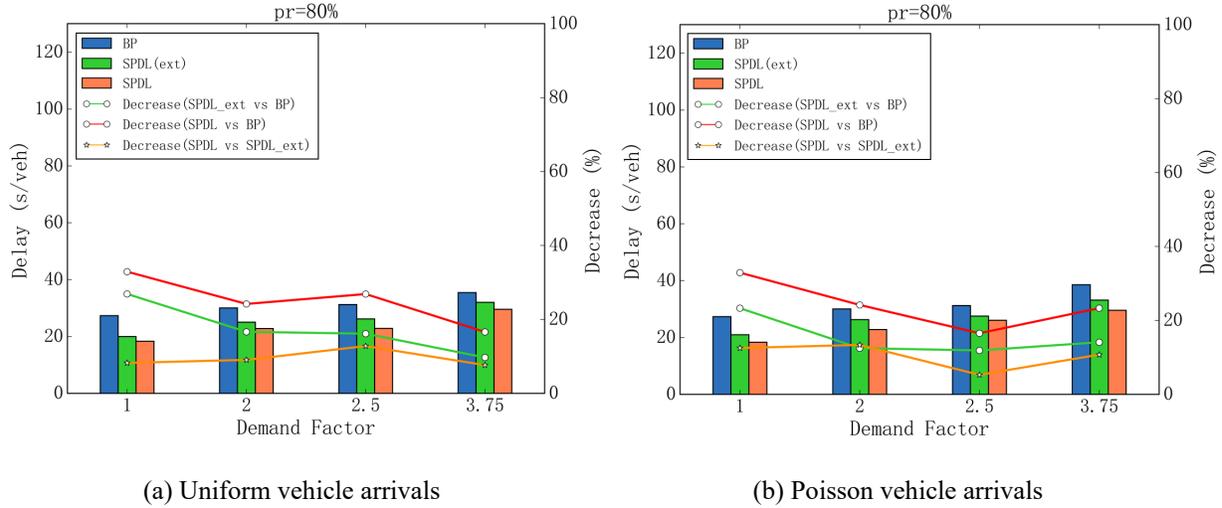

(a) Uniform vehicle arrivals  (b) Poisson vehicle arrivals

**Fig. 25.** Average vehicle delays of the extended model with the CAV proportion rate of 80%.

## 6. Conclusions

This study proposes an SPDL-based traffic control model at isolated intersections under the mixed traffic environment of HVs and CAVs. Left-turn and through CAVs share CAV-dedicated lanes and cross the intersection during the shared phases with HVs. A three-level optimization model is formulated to optimize signal timings and CAV trajectories in one framework. At the upper level, a standard NEMA ring barrier structure is used for the signal optimization and barrier durations are optimized by DP. At the middle level, phase sequence and phase durations are optimized by enumeration for the given barrier from the upper level and the minimum vehicle delay is fed to the upper level. At the lower level, CAV platooning in the buffer zone and trajectory planning in the passing zone are conducted based on the signal timings of the barrier from the middle level and the travel time of CAVs is fed to the middle level. The SPDL-based model could be extended to eliminate the setting of the buffer zone when the head-lag left-turn phase sequence is used. A rolling horizon scheme is designed to apply the proposed model dynamically with time-varying traffic conditions. Simulation results validate the advantages of the SPDL-based control over the benchmark BP-based control in terms of both average vehicle delay and intersection capacity. Sensitivity analyses show that: 1) the SPDL-based control significantly outperforms the BP-based control, especially with high traffic demand; 2) the performance of the SPDL-based control is robust to CAV penetration rates; and 3) the SPDL-based control performs better with balanced left-turn and through traffic demand. In addition, the extended SPDL-based control without the buffer zone could serve as an alternative approach when it is difficult to set the buffer zone.

In this study, only longitudinal acceleration profiles are considered in the trajectory planning for CAVs. In the future research, the lateral trajectories, i.e., lane-changing maneuvers, are expected to be taken into consideration, especially with multiple CAV-dedicated lanes. It is assumed that CAV platooning could be finished in the buffer zone. However, the detailed modeling of CAV platooning and the interactions with HVs is a great challenge, which is worth investigation. Further, the extension of the proposed SPDL-based model to the network level is another research direction.

**Acknowledgements**

This study is supported by National Key Research and Development Program of China (No. 2016YFE0206800), National Natural Science Foundation of China (No. 61903276), and "Shuguang Program" supported by Shanghai Education Development Foundation and Shanghai Municipal Education Commission (No.19SG16).

**Appendix A**

The trajectory planning model (**P3**) is a nonlinear programming model with a hierarchical multi-objective optimization structure. It is solved by the following steps.

**Step 1: Determine the feasible region of the passing time $t_f^\omega$ considering Eqs. (27)-(36)**

Given travel distance $L$, maximum speed $v_{max}$, maximum acceleration rate $a^U$, and maximum deceleration rate $a^L$, the upper bound $t_f^U$ and the lower bound $t_f^L$ of the passing time $t_f^\omega$ can be easily identified. **Fig. 26** shows the three cases according to the relationship between the initial speed $v_0^\omega$ and the critical speeds (i.e., $\sqrt{v_{max}^2 - 2a^U L}$ and $\sqrt{2a^L L}$) when constraints Eqs. (27)-(36) are taken into consideration. The lower bound $t_f^L$ is determined as

$$t_f^L = \begin{cases} t_0^\omega + \frac{v_{max} - v_0^\omega}{a^U} + \frac{L}{v_{max}} - \frac{v_{max}^2 - (v_0^\omega)^2}{2a^U v_{max}}, & \text{if } \sqrt{v_{max}^2 - 2a^U L} < v_0^\omega \leq v_{max} \\ t_0^\omega + \frac{\sqrt{(v_0^\omega)^2 + 2a^U L} - v_0^\omega}{a^U}, & \text{if } 0 \leq v_0^\omega \leq \sqrt{v_{max}^2 - 2a^U L} \end{cases} \quad (43)$$

The upper bound $t_f^U$ is determined as

$$t_f^U = \begin{cases} t_0^\omega + \dfrac{v_0^\omega - \sqrt{(v_0^\omega)^2 - 2a^L L}}{a^L}, & \text{if } \sqrt{2a^L L} \leq v_0^\omega \leq v_{max} \\ +\infty, & \text{if } 0 \leq v_0^\omega \leq \sqrt{2a^L L} \end{cases} \tag{44}$$

Therefore, the feasible region of $t_f^\omega$ is determined considering constraints Eqs. (27)-(36):

$$t_f^L \leq t_f^\omega \leq t_f^U \tag{45}$$

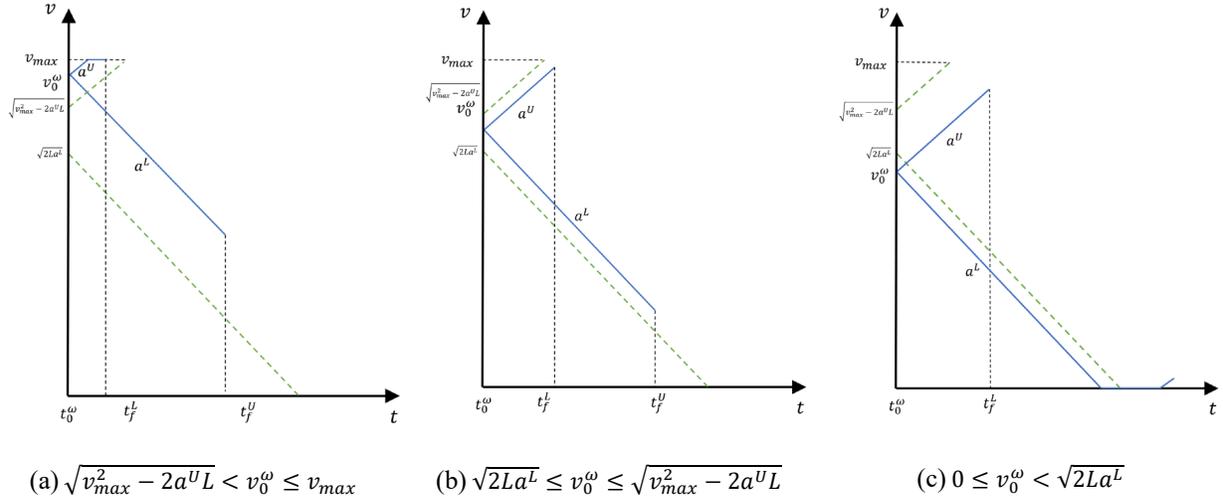

(a) $\sqrt{v_{max}^2 - 2a^U L} < v_0^\omega \leq v_{max}$  (b) $\sqrt{2La^L} \leq v_0^\omega \leq \sqrt{v_{max}^2 - 2a^U L}$  (c) $0 \leq v_0^\omega < \sqrt{2La^L}$

**Fig. 26** Feasible region of $t_f^\omega$ considering Eqs. (27)-(36).

**Step 2: Identify the value of the primary objective function**

Considering constraints Eq. (37) and Eq. (45), the feasible region of $t_f^\omega$ in problem (**P3**) is

$$\max\left(t_{p,r,j}^g, t_f^L\right) \leq t_f^\omega \leq \min\left(t_{p,r,j}^g + g_{p,r,j}\Delta t, t_f^U\right) \tag{46}$$

Since minimizing $t_f^\omega$ is the primary objective, the minimum value of $t_f^\omega$ can be easily identified based on Eq. (46).

*Case 1:* $t_f^L > t_{p,r,j}^g + g_{p,r,j}\Delta t$. The trajectory planning model (**P3**) is infeasible.

*Case 2:* $t_{p,r,j}^g \leq t_f^L \leq t_{p,r,j}^g + g_{p,r,j}\Delta t$. The minimum value of $t_f^\omega$ is $t_f^L$, i.e., $\min t_f^\omega = t_f^L$.

*Case 3:* $t_f^L < t_{p,r,j}^g \leq t_f^U$. The minimum value of $t_f^\omega$ is $t_{p,r,j}^g$, i.e., $\min t_f^\omega = t_{p,r,j}^g$.

*Case 4:* $t_f^U < t_{p,r,j}^g$. The trajectory planning model (**P3**) is infeasible.

**Step 3: Find solutions to problem (P3)**

*Case 1:* $t_f^L > t_{p,r,j}^g + g_{p,r,j}\Delta t$ & *Case 4:* $t_f^U < t_{p,r,j}^g$

Model (**P3**) is infeasible. That is, CAV $\omega$ cannot cross the intersection during the corresponding phase $p$ in the barrier given by the middle level model.

*Case 2:* $t_{p,r,j}^g \le t_f^L \le t_{p,r,j}^g + g_{p,r,j}\Delta t$

When $t_f^\omega = t_f^L$, the number of trajectory segments can be identified as well as the corresponding acceleration/deceleration rate $a_i^\omega$ on each trajectory segment according to **Fig. 26**. For example, when $\sqrt{v_{max}^2 - 2a^U L} \le v_0^\omega \le v_{max}$, **Fig. 26(a)** shows two trajectory segments are needed. On the 1st trajectory segment, CAV $\omega$ accelerates with $a^U$ until the maximum speed $v_{max}$ is reached. On the 2nd trajectory segment, CAV $\omega$ travels with $v_{max}$ and pass the stop line at time $t_f^L$. The end times $t_i^\omega$ of the trajectory segments are then easily calculated by solving equations Eqs. (27)-(28), (31)-(36). And the unique solution can be found. Note that the final speed $v^\omega(t_f^\omega)$ is unique as well in this case.

*Case 3:* $t_f^L < t_{p,r,j}^g \le t_f^U$

When $t_f^\omega = t_{p,r,j}^g$, enumeration is used to find the solutions. There are three trajectory segments and each segment could take the acceleration/deceleration rate of $a^U$, 0, or $a^L$. Therefore, there are nine possible solutions. Each solution can be found by solving Eqs. (27)-(28), (31)-(36) fixing the acceleration/deceleration rate $a_i^\omega$ on each trajectory segment. Among the nine possible solutions, the one with the largest final speed $v^\omega(t_f^\omega)$ is selected as the solution to model (**P3**).